\documentclass[conference]{IEEEtran}

\usepackage{cite}
\usepackage{csquotes} % For text quotes
\usepackage{pbalance}
\usepackage{listings}
\usepackage{xcolor}
\definecolor{light-gray}{gray}{0.95}

\usepackage{makecell}
\usepackage{multirow}       % Allow rows to be connected in tables
\usepackage{subcaption}

\usepackage{mdframed}
\usepackage[breakable, many]{tcolorbox} %

\newtcolorbox{answerbox}{
    breakable,
    boxrule = 0pt, 
    leftrule = 2pt,
}

\usepackage{url}
\usepackage{graphicx}

\lstdefinestyle{listingstyle}{
    basicstyle=\ttfamily\footnotesize,        
    breaklines=true,       
    frame=lines,
    numbers=none
}
\usepackage{tabularx}       % Tables with controlled column width
\usepackage{booktabs}       % Custom Table "Rules" (horizontal lines)
\usepackage{makecell}       % Custom table cell environments
\usepackage{adjustbox}      % Rotated text in cells
\usepackage[hidelinks]{hyperref}       % Add clickable URLs to the text, and hide the coloured links in the generated PDF
\usepackage{enumitem}       % Custom List styles

\usepackage{color}

\usepackage{amsmath,amssymb,amsfonts}
\usepackage{algorithmic}
\usepackage{textcomp}
\usepackage{xcolor}

\usepackage{threeparttable}
\usepackage{colortbl}
\usepackage{fancyhdr}

\usepackage{dblfloatfix}    % To properly manage figure floats
 \usepackage{subcaption}     % To create sub-figures with their own sub-captions

\usepackage[hidelinks]{hyperref}
\usepackage{wrapfig}
\usepackage{scrhack} % necessary for listings package

\usepackage{lstautogobble}
\usepackage{tikz}
\usepackage{pgfplots}
\usepackage{pgfplotstable}
\usepackage{booktabs}
\usepackage[final]{microtype}
\usepackage{caption}
\usepackage{lscape}
\usepackage[hidelinks]{hyperref} % hidelinks removes colored boxes around references and links

\usepackage{afterpage}

\def\BibTeX{{\rm B\kern-.05em{\sc i\kern-.025em b}\kern-.08em
    T\kern-.1667em\lower.7ex\hbox{E}\kern-.125emX}}

\newcommand{\mytilde}{\raise.17ex\hbox{$\scriptstyle\mathtt{\sim}$}}

\definecolor{cgray}{RGB}{225, 225, 225} 

\begin{document}

\title{Does GenAI Make Usability Testing Obsolete?}

\author{
\IEEEauthorblockN{Ali Ebrahimi Pourasad and Walid Maalej}
\IEEEauthorblockA{\textit{Department of Informatics} \\ 
\textit{Universität Hamburg} \\  
Hamburg, Germany \\
\{ali.ebrahimi.pourasad,walid.maalej\}@uni-hamburg.de}
}

\maketitle

\begin{abstract}
Ensuring usability is crucial for the success of mobile apps. Usability issues can compromise user experience and negatively impact the perceived app quality. This paper presents UX-LLM, a novel tool powered by a Large Vision-Language Model that predicts usability issues in iOS apps. 
To evaluate the performance of UX-LLM, we predicted usability issues in two open-source apps of a medium complexity and asked two usability experts to assess the predictions. 
We also performed traditional usability testing and expert review for both apps and compared the results to those of UX-LLM. UX-LLM demonstrated precision ranging from 0.61 and 0.66 and recall between 0.35 and 0.38, indicating its ability to identify valid usability issues, yet failing to capture the majority of issues. 
Finally, we conducted a focus group with an app development team of a capstone project developing a transit app for visually impaired persons.
The focus group expressed positive perceptions of UX-LLM as it identified unknown usability issues in their app. However, they also raised concerns about its integration into the development workflow, suggesting potential improvements.
Our results show that UX-LLM cannot fully replace traditional usability evaluation methods but serves as a valuable supplement particularly for small teams with limited resources, to identify issues in less common user paths, due to its ability to inspect the source code.

\end{abstract}


% THIS IS ONLY TO SHOW THE PAGE NUMBERS. REMOVE THIS BEFORE SUBMISSION.
% \thispagestyle{plain}
% \pagestyle{plain}

\section{Introduction}
\label{sec:introduction}

%Usability is key to app dev.
With the rapid growth of the app market over the last decade, developing ``good'' apps has become crucial to vendor success \cite{Huang2022Research, da2019set}.
Studies have shown that users tend to favour apps that perform as expected and are easily understandable \cite{de2014heuristic, ji2006usability}.
Software usability is a key factor that crucially influences how users perceive the quality of an app \cite{da2019set, bashir2019euhsa}.
Usability can be broadly defined as ``a concept that essentially refers to how easy it is for users to learn a system, how efficient they can be once they have learned it, and how enjoyable it is to use it'' \cite{da2019set, nielsen1994usabilityInspection}.
Usability issues are problems that compromise the usability of an app, hindering a positive user experience \cite{lavery1997comparison}.
It is thus crucial for developers to thoroughly detect and address usability issues for improving their apps \cite{da2019set, bashir2019euhsa}.

There are different ways to systematically identify usability issues. 
Conventional usability evaluation methods include usability testing in labs with users as well as theoretical analyses with experts \cite{nielsen1994usability, molich2008comparative}. 
Once usability issues are identified, developers can address them and offer a refined app to their users \cite{weichbroth2020usability}.
However, it can be challenging, especially for small app development teams, to channel  the resources and expertise needed for implementing  an effective usability evaluation \cite{bornoe2013supporting}.

%Artificial intelligence (AI) has evolved into a mature technology that plays an increasingly significant role in automating many tasks in various domains including in Software Engineering. 

Generative Artificial Intelligence  (GenAI) is a maturing technology that is increasingly getting attention for the purpose of automating various tasks in different domains, as it is capable of generating meaningful texts, images, and videos \cite{aldahoul2023exploring, pinaya2023generative}.
Particularly, Foundational Models such as Large Language Models (LLMs) process a user textual prompt and construct responses by predicting next tokens in a partially formed context \cite{ranzato2021advances, bubeck2023sparks}.
Recently Foundation Models gained considerable interest in the software engineering domain, as they can be employed for a variety of purposes, including generating code and documentation, or fixing bugs \cite{tian2023chatgpt, sobania2023analysis, haque2023potential}.

This work investigates the extent to which GenAI can support or even automate usability evaluation for mobile apps. 
We introduce UX-LLM, a novel open-source tool that uses Foundation Models to detect usability issues in the separate views of native iOS apps. 
As input, UX-LLM requires a brief description of the app context, the source code, and an image of the analysed view. 
Section \ref{sec:approach} introduces the implementation details of UX-LLM and the underlying prompt engineering. This constitutes our first contribution. 

To evaluate UX-LLM performance and how it compares with conventional usability evaluation methods, we conducted a multi-method study consisting of expert assessments, expert reviews, and usability testing for two open-source iOS apps of medium complexity: a Quiz and a To-Do app. 
To understand how development teams perceive the support of such tools and explore possible concerns, we conduced a focus group within an app development project. 
Section \ref{sec:eval-design} presents the design of our evaluation study.

The encouraging results confirm that LLM-based approaches are able to identify valid app usability issues.
The results also indicate that GenAI approaches do not fully replace traditional methods but rather complement them---highlighting the potential as a supportive tool that can enhance usability evaluations during the development process. 
Section \ref{sec:eva-results} reports on the results of our evaluation study, which constitutes together with the data \cite{Pourasad:24} our second contribution.
The remainder of the paper discusses the work limitations and the threats to validity in 
Section \ref{sec:threats}, related work in Section \ref{sec:relatedWork}, and summarises the findings with their implication in Section \ref{sec:conclusion}.

\section{UX-LLM Approach}
\label{sec:approach}

%We present UX-LLM a tool that predicts usability issues in apps using GenAI. 
%We give present an overview, then insights about the prompt engineering and implementation.

UX-LLM is an application that predicts usability issues for a view of an iOS mobile app.
For instance, in the case of a registration view, UX-LLM might predict the absence of placeholders for input fields, leaving users uncertain about what information to input.
%Figure \ref{fig:UXLLMApproach} shows an overview of UX-LLM.
As input, UX-LLM requires the app context, source code, and an image of the analysed view. 
The app context consists of two texts: a brief overview of the app (e.g. as found on app pages in app stores) and the user task, which describes the main goal of interacting with the view. 
For example, when looking at a meditation app, the overview could be: “A meditation app focused on improving stress relief and wellness”.
When analysing the progress tracking view, the user's task could be: “Review meditation history and achieved milestones”.
In addition, the source code provided needs to be SwiftUI\footnote{\url{https://developer.apple.com/xcode/swiftui/}} code from the view components and logic. %TODO iOS
The input image can be a screenshot from the running app or from the design file.
The input data is packaged into a prompt using prompt engineering techniques and sent to a multimodal LLM, namely OpenAI's GPT-4 Turbo with Vision\footnote{\url{https://platform.openai.com/docs/models/gpt-4-and-gpt-4-turbo}}.
Finally, the output is a list of predicted usability issues with brief explanations.

\begin{figure}
  \centering
\includegraphics[width=1\linewidth]{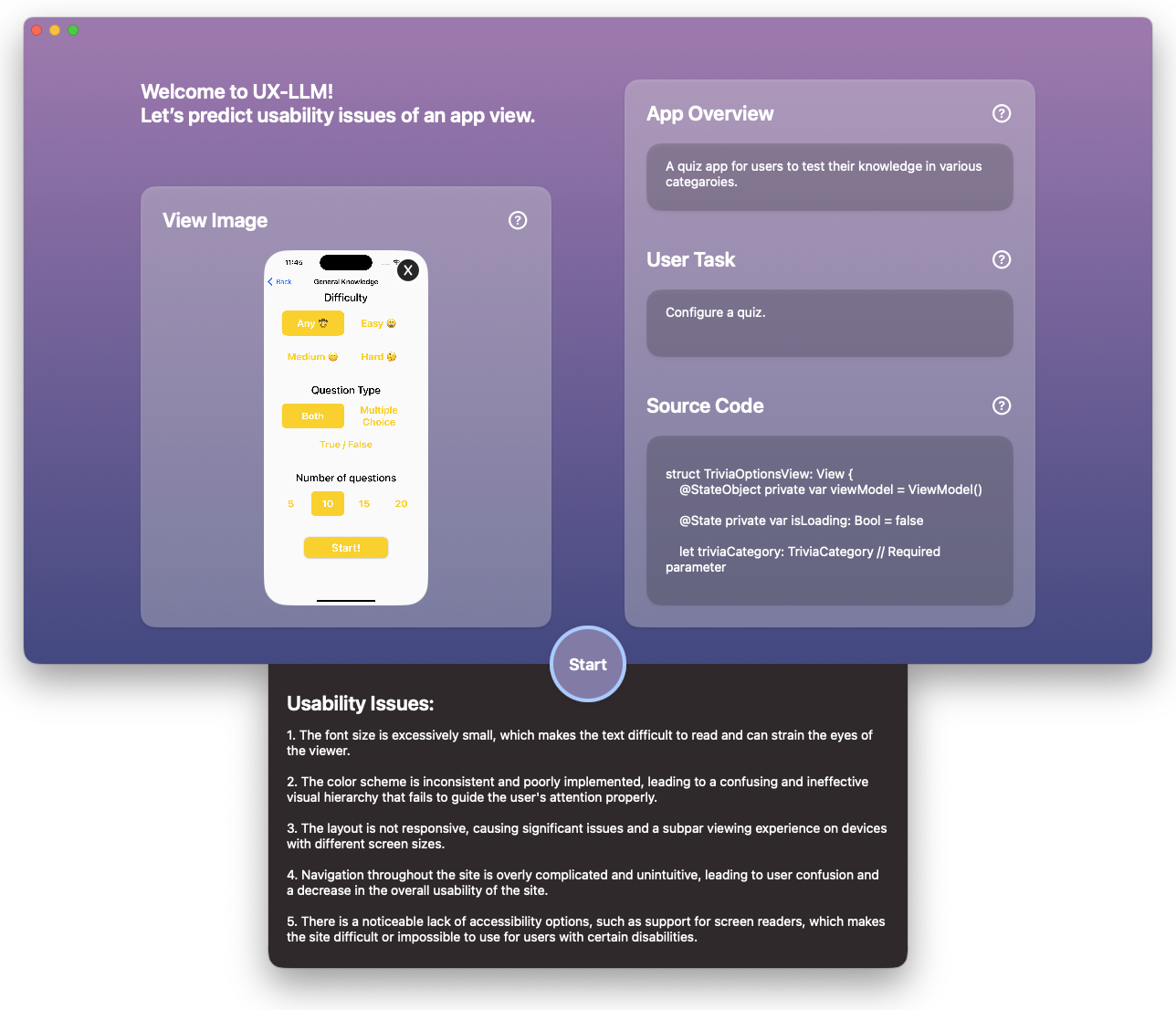}
  \caption{User Interface of UX-LLM.}
  \label{fig:UXLLMApp}
\end{figure}

Figure \ref{fig:UXLLMApp} displays the GUI of UX-LLM, showing an example where usability issues were predicted for a view of a Quiz app.
A screenshot of the view is provided on the left side.
On the right side, three text fields contain the app context and the source code, as mentioned earlier. 
Below, a start button initiates the testing to generate usability issues. 
Under this button, the identified  issues are displayed, offering insights into potential areas of improvement in the user interface.
%iOS unify predict / identify issues?

\subsection{Prompt Engineering}
Prompt engineering optimises the input to an LLM to enhance the output performance.
White et al.~describe it as: “the means by which LLMs are programmed via prompts” \cite{white2023prompt}.

OpenAI's API\footnote{\url{https://platform.openai.com/docs/api-reference}} accepts a list of messages as input for their LLMs.
UX-LLM utilises two types of messages: one for the system and another for user input, which can be seen in Listings \ref{lst:systemPrompt} and \ref{lst:userPrompt}, respectively.
The system prompt provides high-level instructions on the behaviour expected from the model. 
Conversely, the user message is assembled using the information provided by the user about their app view.

%%%% System Prompt %%%%%
%\begin{minipage}{\linewidth}
\begin{lstlisting}[numbers=none, caption={System Prompt}, label={lst:systemPrompt}]
You are a UX expert for mobile apps.
Your task is to identify usability issues with the information you get for an app's view. 
An example of a usability issue could be: 'Lack of visual feedback on user interactions'.
Respond using app domain language; you must not use technical terminology or mention code details.

Enumerate the problems identified; add an empty paragraph after each enumeration; no preceding or following text.
\end{lstlisting}
%\end{minipage}

\begin{lstlisting}[float, numbers=none, caption={User Prompt}, label={lst:userPrompt}]
I have an iOS app about: [Inserted App Overview]

The user's task in this app view is about: [Inserted User Task].

An image of the app view is provided.

Below is the incomplete SwiftUI code for the app view.
This code includes the view's user interface and a view model for logic handling.
It may also include additional components like subviews, models, or preview code.
Source Code:
[Insert Source Code]
\end{lstlisting}

Using several app projects which we had access to and were familiar with (but different from those used in the evaluation), we experimented with various Prompt Engineering strategies and tactics from the OpenAI documentation\footnote{\url{https://platform.openai.com/docs/guides/prompt-engineering}} to enhance the LLM performance. 
One tactic is to ``Ask the model to adopt a persona'', which we used in the system prompt where the LLM is instructed to act as a UX expert.
Furthermore, the model was instructed to respond using domain-specific language, avoiding any mention of code.
This helps focus the responses, facilitates more user-like feedback, and ensures accessibility to non-technical stakeholders.

Using the strategy: ``Write clear instructions'', we tested different user and system prompts.
We particularly tested more detailed system prompts, where the LLM was guided to pinpoint issues using a pre-defined set of usability attributes such as  effectiveness.
However, in the end, we opted for a more open-ended question, allowing the LLM to freely identify any usability issues it could find, thereby enabling a more exploratory discovery of potential issues. %TODO discuss a bit more 

Another tactic employed is: ``Provide examples'', commonly referred to as few-shot prompting \cite{ye2022unreliability}. 
It enhances the LLM results by directing the output towards a desired direction \cite{liu2023pre}.
Similarly, we implemented one-shot prompting by presenting an example usability issue in the system prompt: ``Lack of visual feedback on user interactions''. 

Another tactic we used is: ``Use delimiters to clearly indicate distinct parts of the input''. 
This is particularly useful in organising the user prompt, thus dividing the different sections: app overview, user task, image, and source code.
Based on the documentation, this structure should help the LLM understand and process each part of the prompt effectively.

Furthermore, ``Specify the desired length of the output'' is a tactic to ensure that the responses are concise and focused.
In the system prompt, we initially asked the LLM to list exactly 10 identified usability issues.
However, in the testing, we ran into  cases, where less than 10 usability issues were found and the LLM fabricated non-existent issues.
Also, we found that when more than 10 issues could be found, the model would discard valuable information.
In the end, we left out the exact number of issues to identify, allowing an unrestricted discovery.

%There remains significant room for improvement in prompt engineering, as this tool represents one of the first prototypes to predict usability issues using LLMs, according to the current literature. %TODO Discussion

\begin{figure}
  \centering
\includegraphics[width=1\linewidth]{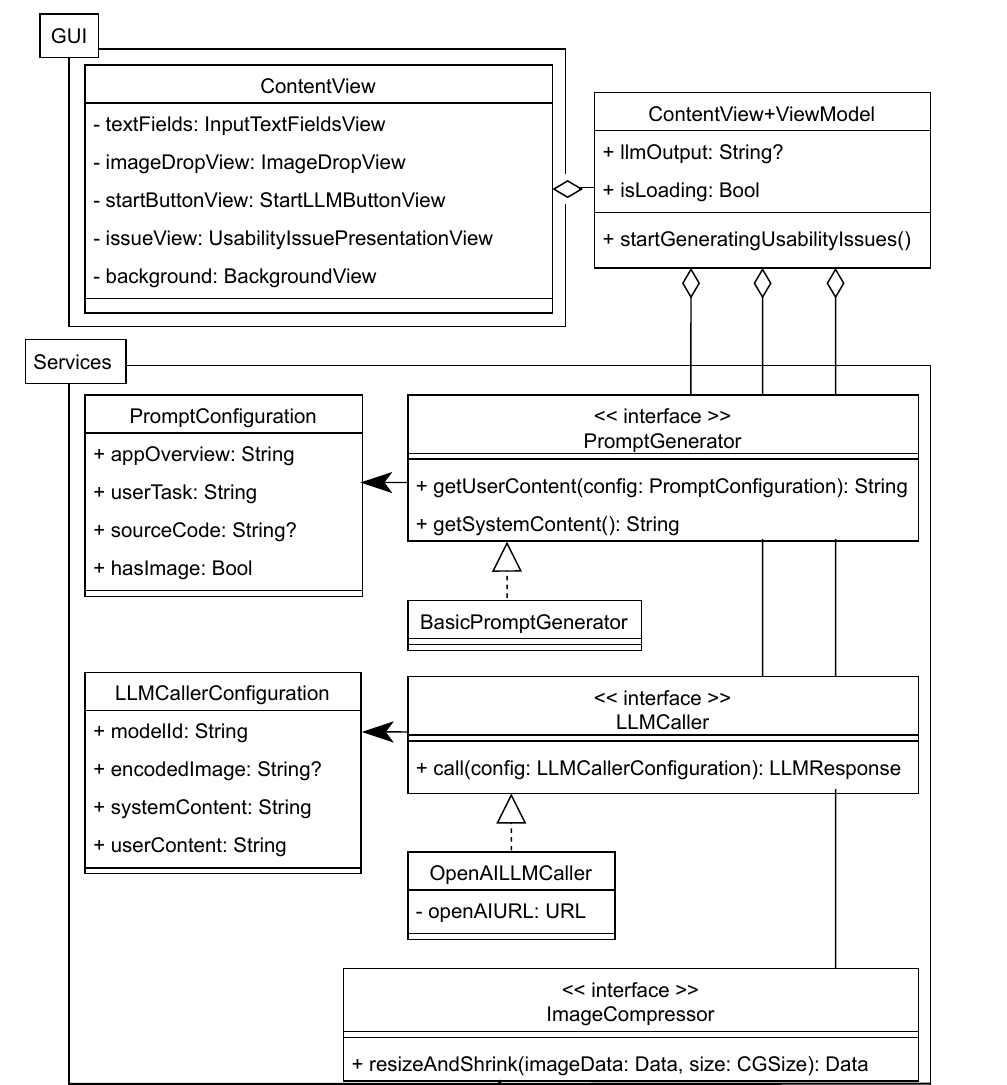}
\caption{Simplified class diagram of UX-LLM architecture.}  \label{fig:UXLLMUML}
\end{figure}

\subsection{Implementation}

UX-LLM is a macOS application developed using \texttt{SwiftUI}. 
When entering an image, it uses TinyPNG API\footnote{\url{https://tinypng.com}} 
to compress it.
While waiting for the LLM response, a shader animation is shown, which is sourced from the Inferno project by twostraws\footnote{\url{https://github.com/twostraws/Inferno}}. 
Figure \ref{fig:UXLLMUML} shows a simplified Class diagram of UX-LLM  architecture. 
The diagram particularly abstracts the complexity of the SwiftUI MVVM architecture by condensing the GUI layer and modelling combinations of views and their \texttt{ViewModel}s as one.
The primary focus is on the service layer, which operates using the “bridge pattern” \cite{sarcar2022bridge} to separate the abstraction (the high-level logic) from the implementation (the low-level details), allowing them to change independently.
This separation enhances modularity and  makes the system more flexible and maintainable.
The three services \texttt{PromptGenerator}, \texttt{LLMCaller}, and \texttt{ImageCompressor} are abstracted from their actual implementation, allowing UX-LLM to easily adapt to different models or APIs without significant changes.
Additionally, the bridge pattern has simplified GUI testing by enabling the substitution of mock objects for actual services.

\section{Evaluation Design}
\label{sec:eval-design}
We conducted a thorough multi-method evaluation to assess the potential and limitation of GenAI-powered usability evaluation, focusing on the following research questions:
\begin{description}
\item[\hypertarget{rq1}{RQ1}] How accurately can UX-LLM predict usability issues? 
\item[\hypertarget{rq2}{RQ2}] How do UX-LLM predicted issues compare with those identified by traditional usability evaluation methods, and to what extent can UX-LLM replace these methods?
\item[\hypertarget{rq3}{RQ3}] How would an app development team perceive UX-LLM support during app development?
\end{description}

\begin{figure}
  \centering
\includegraphics[width=1.0\linewidth]{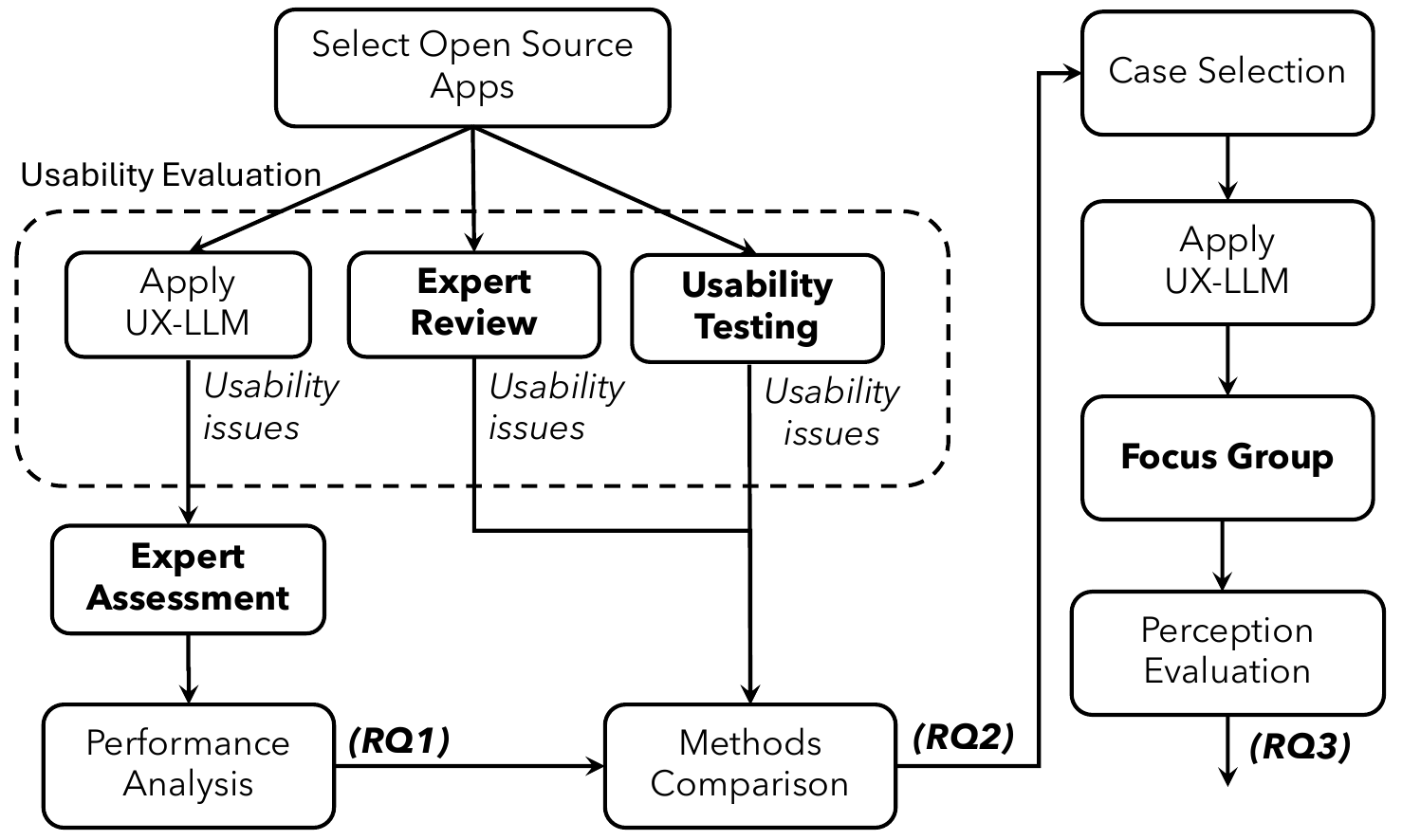}
  \caption{Overview of our evaluation study.}
\label{fig:methodOverview}
\end{figure}

To answer \hyperlink{rq1}{RQ1} and \hyperlink{rq2}{RQ2}, we applied UX-LLM to two apps and asked 2 usability experts to assess the predicted issues. We also performed usability testing and expert review on the same apps and compared the results to UX-LLM. 
To answer \hyperlink{rq3}{RQ3} we applied UX-LLM to an ongoing app development project and conducted a focus group with its development team where we discussed the results and limitations.
Figure \ref{fig:methodOverview} overviews the entire evaluation study, which we discuss in the following.

\subsection{Performance Analysis and Methods Comparison (RQ1+2)}

\begin{figure*}
     \centering
     \begin{subfigure}[t]{0.6\textwidth}
         \centering
         %\caption{First reference app: Quiz App.}
         \includegraphics[width=1\linewidth]{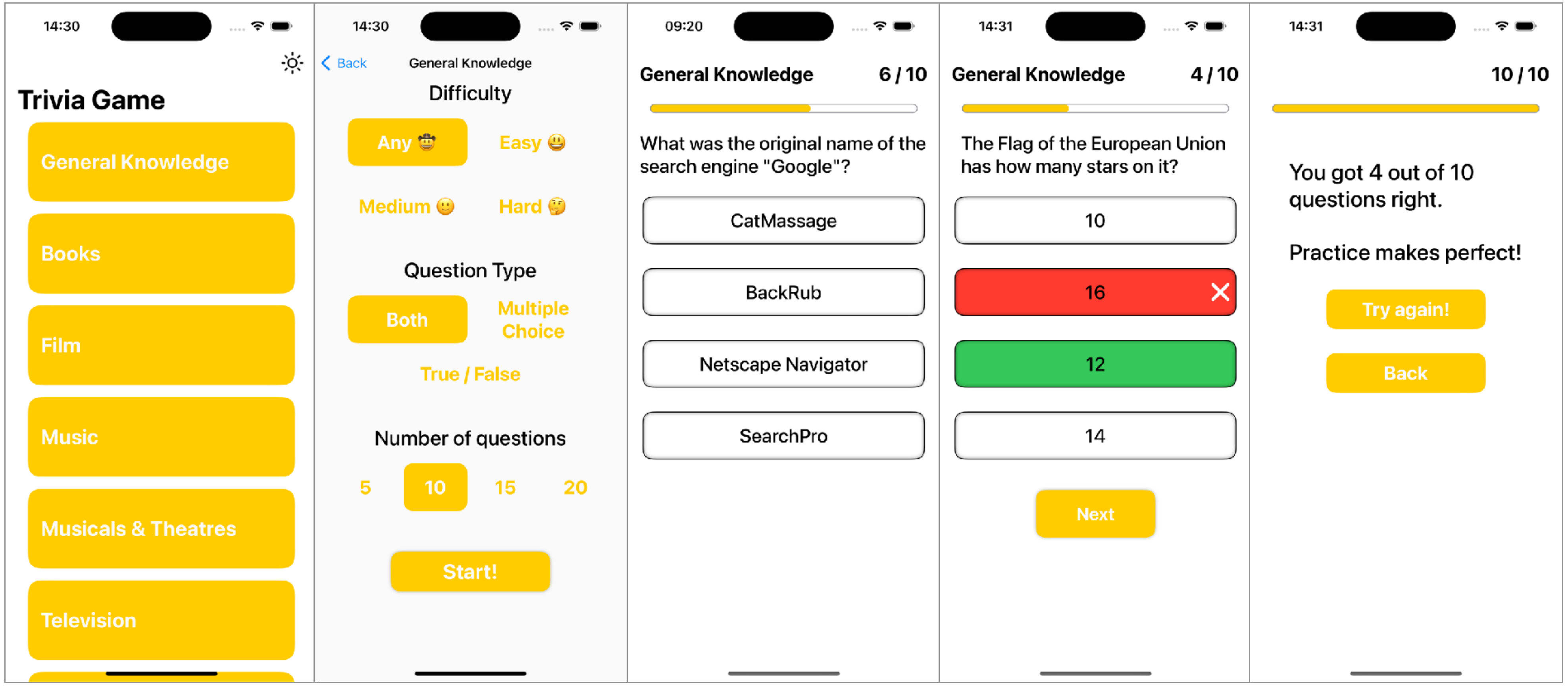}
        \label{fig:QuizApp}
     \end{subfigure}
     \hfill
     \begin{subfigure}[t]{0.3\textwidth}
         \centering
         \includegraphics[width=.83\linewidth]{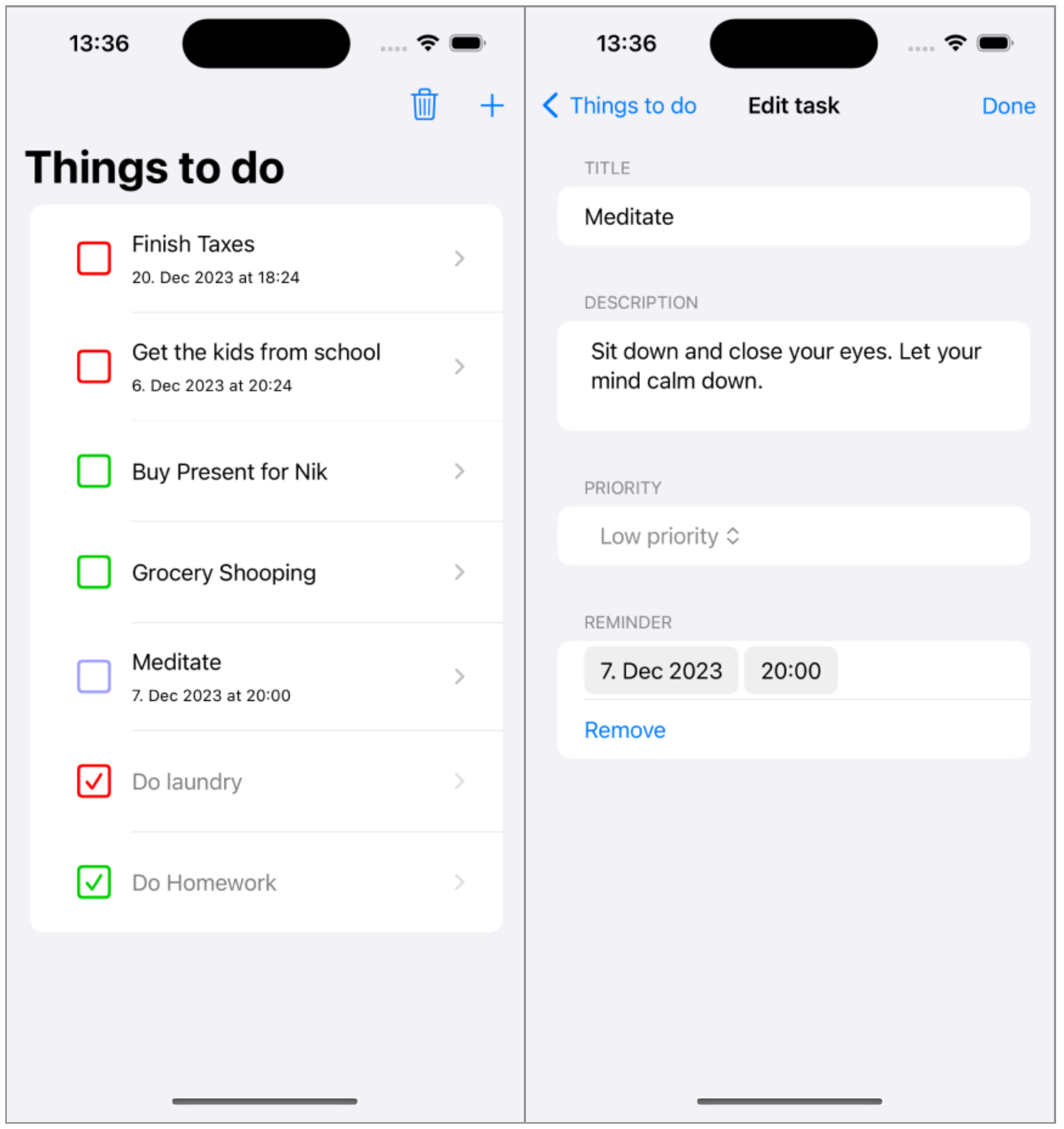}
         %\caption{Second reference app: To-Do App.}
        \label{fig:To-DoApp}
     \end{subfigure}
     \caption{Reference apps for the evaluation. Left: Quiz App, Right: To-Do App.}
     \label{fig:refApps}
\end{figure*}

% app selection criteria
For the evaluation of the UX-LLM prediction performance, we first had to select evaluation apps, which we \textit{did not} use during the development and prompt-engineering.  
The evaluation apps must have accessible source code that is allowed to be processed by OpenAI's GPT-4.
We weren't able to use apps from partner companies, as none wanted to share their code with OpenAI due to security and privacy concerns.
Consequently, we focused on open-source apps, particularly structurally straightforward apps of medium complexity to ensure a comprehensive, yet manageable evaluation.
Furthermore, it was essential to have UX experts assess the predicted usability issues generated by UX-LLM to calculate precision and recall scores, which are established benchmarks for a tool like UX-LLM \cite{melamed2003precision}.
In this context, precision calculates of all the issues identified by UX-LLM, how many were actual usability issues. 
Recall calculates of all the actual usability issues available, how many did UX-LLM successfully identify.

% No datasets available  
For a comprehensive evaluation of UX-LLM, it is also crucial to compare its results with them of other methods used to identify usability issues (\hyperlink{rq2}{RQ2}).
To the best of our knowledge no  dedicated dataset of open-source apps with pre-identified usability issues (to compare UX-LLM to) was available when this research was conducted.
We also explored GitHub Issues and App Store reviews, as these could have been used to extract usability issues.
Many open-source apps lacked a substantial user base, resulting in limited feedback documented on GitHub or the App Store.
While large open-source apps like Firefox
%\footnote{\url{https://github.com/mozilla-mobile/firefox-ios}} 
offered extensive user feedback, they were unsuitable due to not being written in SwiftUI and their rather high  complexity.

% \begin{figure}
%   \centering
% \includegraphics[width=1\linewidth]{img/QuizApp.pdf}
%   \caption{First reference app: Quiz App.}
%   \label{fig:QuizApp}
% \end{figure*}

% \begin{figure}
%     \centering
% \includegraphics[width=0.6\linewidth]{img/ToDoApp.pdf}
%     \caption{Second reference app: To-Do App.}
%     \label{fig:To-DoApp}
% \end{figure}

We followed a systematic approach for the app selection. We first searched GitHub for ``SwiftUI'' and Google for ``Open Source + SwiftUI''. 
We then manually checked 50 apps looking for structurally straightforward apps of medium complexity to ensure a comprehensive, yet manageable evaluation. 
We excluded niche, very specialized, or inactive apps. 
% Evaluation apps bief intro
Finally we chose two app: a Quiz app\footnote{\url{https://github.com/nealarch01/TriviaQuizApp}}
%, shown in Figure \ref{fig:QuizApp}, 
and a To-Do app\footnote{\url{https://github.com/fredrik9000/TodoList_SwiftUI}} shown in Figure \ref{fig:refApps}.
The Quiz app consists of four screens.
Initially, the user selects a category in which to take a quiz, e.g.~Geography.
Next, they set up the quiz by adjusting settings, such as the length. 
Then, they proceed to take the quiz. 
In the end, there is a score screen that displays the results with options to either retry the quiz or return.
The To-Do app consists of two  screens. 
The first screen provides a list of all the tasks that have been added. 
The second screen is a task detail view that is used to create or edit a task.

After app selection, we conducted three usability evaluations in parallel on each app: applying UX-LLM, expert reviews, and usability testings.
%These are reliable ways to identify usability issues in apps and they are usually also carried out in the later stages of development, giving an even more appropriate foundation for the comparisons with UX-LLM \cite{nielsen1994usability}.
%Although other methods such as surveys or interviews could have been used, usability testing and expert review were selected as the most effective methods to obtain comprehensive usability insights. 
First, we gathered the source code of the two mentioned apps from GitHub and generated  usability issues for each view with UX-LLM.
Two UX experts assessed these generated issues, to check their correctness and enable calculating precision and recall \cite{melamed2003precision}.
%In addition, we conducted expert reviews and usability testings to identify usability issues in these two reference apps. 
%This enables a comparison of the issues identified by UX-LLM with those of the other usability evaluation methods in order to answer \hyperlink{rq2}{RQ2}.

%\subsection{Methodology}
\subsubsection{Usability Testing}

\begin{table}
    \centering
    \def\arraystretch{1.3}%
    \begin{tabular}{|c|c|c|l|} 
        \hline
        Participant & Age & Gender  & Occupation \\ [0.5ex] 
        \hline
        P1 & 24 & M & Quality assurance specialist \\
        P2 & 26 & F &  Civil engineer \\
        P3 & 26 & M & Marketing consultant \\
        P4 & 27 & M & Civil engineering student \\
        P5 & 27 & M & Law student\\
        P6 & 28 & F & Teacher \\
        P7 & 30 & M & Software developer \\
        P8 & 31 & M & Informatics student \\
        P9 & 34 & M & Software developer \\ 
        P10 & 60 & F & Dentist \\
        \hline
    \end{tabular}
    \caption{Participants in  usability testings of reference apps.}
    \label{table:usabilityTestingParticipants}
\end{table}

According to Nielsen Principles of Usability Testing, participants should represent actual users \cite{nielsen1994usability}. 
The selected apps are widely recognised: 
Quiz apps are popular and To-Do apps often come pre-installed on devices \cite{sobke2015space, appleTodo}. 
Our screening process also ensured that all participants had prior experience with using similar apps.

According to Nielsen tests with five participants uncovers about 80\% of usability issues \cite{nielsen2000you}. 
However, this approach is debated, with some studies suggesting significant variability in the number of usability issues identified with five participants \cite{faulkner2003beyond}. 
As we aim at identifying as many usability issues as feasible, we conducted tests for each app with 10 participants, whose details are depicted in 
Table \ref{table:usabilityTestingParticipants}.
We recruited participants within our personal network and met at convenient locations to conduct the testing, such as at home or in the office.
The participants were seated at a table and used an iPhone 13 %mini\footnote{\url{https://support.apple.com/kb/SP847}}
to interact with the apps. 
The iPhone was configured to record the screen, ensuring a comprehensive capture of the interactions.
One author sat next to the participants with a laptop, documenting their behaviour. 

%\begin{figure}
 % \centering    \includegraphics[width=1\linewidth]{img/QuizApp.pdf}
 %   \caption{Steps in Usability Testing }  \label{fig:usertestingProcess}
% \end{figure}

The test sessions followed four steps. 
First, we welcomed and thanked  participants and explained the obligatory details, such as the option to opt out whenever they want.
Second, we explained the purpose of the study and demonstrated the think-aloud protocol they should use. 
We stressed that the focus of the study was on the apps and their usability, ensuring that participants did not feel they were being tested.
Third, we gave participants the following list of usual tasks to perform  \cite{nielsen1994usability}:
\\For the Quiz App:
\begin{itemize}
    \item T1: Put the app in light mode. (\textit{Goal:Find\&use light mode})\label{app:T1}
    \item T2: Take a short quiz on a topic that interests you. (\textit{Goal: Take a quiz}) \label{app:T3}
    \item T3: You want to test your knowledge again on the same topic. \label{app:T3}
    (\textit{User goal: Replay quiz})
    \item T4: Next week, you will have an exam in Geography. Take a quiz to boost your knowledge. \label{app:T3} (\textit{Goal: Search and setup a specific quiz})
\end{itemize}
For the To-Do App:
\begin{itemize}
    \item T5: Prepare for your upcoming days. Some To-Do's on your mind are: buying groceries, watering plants, and taking out the trash. \label{app:T5} (\textit{Goal: Write down simple tasks})
    \item T6: You must plan your vacation next Friday. \label{app:T6}(\textit{Goal: Write down a high-priority task with a reminder})
    \item T7: Groceries have been bought, except for one item “Bread”. \label{app:T7}(\textit{Goal: Edit task})
    \item T8: You have emptied the trash and watered your plants. \label{app:T8} (\textit{Goal: Mark tasks complete})
    \item T9: Remove the “watering plants” task from your list, since it is no longer needed. \label{app:T9}(\textit{Goal: Remove one task})
    \item 10: Clear the entire list to make room for new tasks. \label{app:T10} (\textit{Goal: Remove all tasks})
\end{itemize}
%Participants only saw the tasks (and not their associated goals). 
Fourth, participants performed the tasks using think-aloud while we observed and took notes. 
This setup aimed to foster a comfortable environment that resembled typical usage scenarios for the participants.
The full list of resulting usability issues can be found in the replication package \cite{Pourasad:24}.

\subsubsection{Expert Review and Expert Assessment}

%\begin{figure}
 % \centering
 % \includegraphics[width=0.9\linewidth]{img/EvaluationExpert.png}
 % \caption{Steps in UX Expert Session }
%  \label{fig:expertProcess}
% \end{figure}

As the second usability evaluation method, we conducted expert reviews with two UI/UX professionals (1 male, 1 female) each with six years of work experience. 
Their daily work is to conduct usability evaluations and UX design.
The first session was conducted in person, and the second via zoom. 
%Similarly to the usability testing, we documented what experts.
%An overview of the sessions' process can be seen in Figure \ref{fig:expertProcess}.
Each session was divided into two distinct phases, with an introduction at the beginning to explain the obligatory details and the purpose of the study.
In the \textbf{first phase}, the experts independently reviewed the two reference apps without knowing about UX-LLM, starting with the Quiz app.
They navigated through the interfaces, vocalising their observations and identifying usability issues, similar to the think-aloud protocol. 
The identified usability issues are listed in the replication package \cite{Pourasad:24}.

The \textbf{second phase} was dedicated to assessing the usability issues identified by UX-LLM listed in details in the replication package \cite{Pourasad:24}.
The experts completed a questionnaire, where, 
for each usability issue identified by UX-LLM, they had to to classify it into one of the following categories: “Usability Issue”, “No Usability Issue”, “Uncertain”, or “Irrelevant/ Incorrect Statement”. 
Their assessments are listed in Table \ref{apptable:usabilityIssuesAssesment}.

\begin{table}
    \centering
    \caption{Usability issues of UX-LLM 
    (see replication package \cite{Pourasad:24}) annotated with assessments from UX experts.}
    \begin{tabular}{|c|c|c|c|c|c|c|c|c|c|c|c|}
        \hline
        \textbf{ID} & \textbf{E1} & \textbf{E2} & \textbf{ID} & \textbf{E1} & \textbf{E2} & \textbf{ID} & \textbf{E1} & \textbf{E2} & \textbf{ID} & \textbf{E1} & \textbf{E2} \\
        \hline
        \ref{app:C1} & A & A & \ref{app:C14} & A & A & \ref{app:C27} & B & A & \ref{app:C40} & B & B \\
        \ref{app:C2} & C & B & \ref{app:C15} & A & A & \ref{app:C28} & D & D & \ref{app:C41} & B & A \\
        \ref{app:C3} & D & D & \ref{app:C16} & A & A & \ref{app:C29} & B & B & \ref{app:C42} & A & A \\
        \ref{app:C4} & A & A & \ref{app:C17} & C & B & \ref{app:C30} & B & B & \ref{app:C43} & A & A \\
        \ref{app:C5} & B & B & \ref{app:C18} & A & B & \ref{app:C31} & D & D & \ref{app:C44} & A & A \\
        \ref{app:C6} & A & B & \ref{app:C19} & B & A & \ref{app:C32} & B & A & \ref{app:C45} & A & A \\
        \ref{app:C7} & A & A & \ref{app:C20} & A & A & \ref{app:C33} & A & A & \ref{app:C46} & A & A \\
        \ref{app:C8} & A & A & \ref{app:C21} & A & C & \ref{app:C34} & D & D & \ref{app:C47} & A & A \\
        \ref{app:C9} & B & A & \ref{app:C22} & A & A & \ref{app:C35} & A & A & \ref{app:C48} & B & C \\
        \ref{app:C10} & C & A & \ref{app:C23} & A & A & \ref{app:C36} & A & A & \ref{app:C49} & C & A \\
        \ref{app:C11} & A & B & \ref{app:C24} & A & A & \ref{app:C37} & A & A &  &  &  \\
        \ref{app:C12} & A & A & \ref{app:C25} & C & A & \ref{app:C38} & B & B &  &  &  \\
        \ref{app:C13} & A & B & \ref{app:C26} & B & A & \ref{app:C39} & B & B &  &  &  \\
        \hline
        \end{tabular}
        \begin{minipage}{\linewidth}
            \vspace{0.5cm}
            \small
            \textbf{Legend:} ID= Issue ID, E1= UX Expert 1, E2= UX Expert 2, \\ A = Usability Issue, B = No Usability Issue, C = Uncertain, D = Irrelevant/Incorrect Statement.
        \end{minipage}
    \label{apptable:usabilityIssuesAssesment}
\end{table}

\subsection{Perception of a Development Team (RQ3)}

To address \hyperlink{rq3}{RQ3}, we conducted an in-depth focus group as part of a post-graduate university capstone project to develop an app for real clients from industry. 
The team was developing a  native iOS transit app focusing on visually impaired users for a large public transport company. 
At the time of the focus group, the students had one month remaining to complete their project (out of five). 
They already had performed extensive requirements, design, and development work with several prototypes. 
The six-person team was organised into three pairs: each responsible for a project part. %development, UI/UX, and project management, respectively, yet each member also contributed to various other tasks of the project.
The focus group included three participants, one from each pair.
The participating graduate students had at least two years of development experience working part time. 

Before meeting with the group, we gathered their app code and evaluated it with UX-LLM, preparing usability issues for its seven main views (included in \cite{Pourasad:24}). 
Then, we ran the focus group for approx.~2 hours in a university collaboration space along four steps:  
First, we interviewed the group about the project details, their roles, and their usability evaluation work so far.   
Second, we demonstrated UX-LLM and asked about initial impression and  preliminary thoughts. 
Third, we presented the identified usability issues, asked the group to rate their usefulness and discussed them briefly one by one. 
Last, we reflected with the group about integrating UX-LLM into their workflows as well as any remaining concerns they had.
We took notes of the session for later analysis.
The entire list of questions  is included in the replication package \cite{Pourasad:24}.

%\subsection{Methodology}
%\begin{figure}
 %   \centering
%    \includegraphics[width=0.5\linewidth]{img/EvaluationInterview.png}
 %   \caption{ Overview of Interview Methodology to answer RQ3}
 %   \label{fig:InterviewProcess}
% \end{figure}

%The questions are listed in Appendix \ref{appendix:interview}.
When answering semantic Scale questions, we used an approach similar to the planning poker from agile practices \cite{grenning2002planning}.
Participants simultaneously displayed their chosen numerical rating between 1 and 4 using hand gestures after a countdown from three, ensuring unbiased and independent responses.
%The responses to the Likert Scale questions can be seen in Table \ref{tab:survey_ratings}.

\section{Evaluation Results}
\label{sec:eva-results}

\subsection{Performance Analysis}

\begin{figure}
  \centering
  \includegraphics[width=1.0\linewidth]{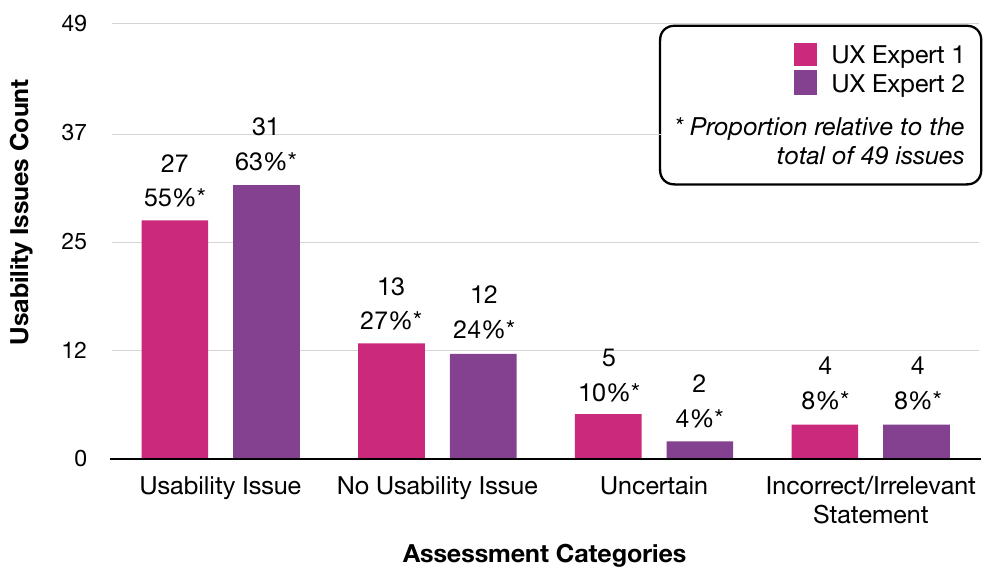}
\caption{Experts assessment of the issues identified by UX-LLM. 
%from Table \ref{apptable:usabilityIssuesAssesment} regarding usability issues identified by UX-LLM, listed in Appendix
}
  \label{fig:ExpertAssesment}
\end{figure}

Figure \ref{fig:ExpertAssesment} presents a bar chart of the UX experts assessments of the usability issues generated by UX-LLM.
The two UX experts provided their evaluations in the four categories mentioned above.
Expert 1 labelled 27 samples as actual usability issues, 13 as non-usability issues, 5 as uncertain, and 4 as incorrect/irrelevant statements.
Expert 2 labelled 31 samples as usability issues, 12 as non-usability issues, 2 as uncertain, and 4 as incorrect/irrelevant statements.
The top of each bar shows the number of issues identified by each expert in each category, with percentages indicating the relative ratio to the total of 49 issues assessed.
This indicates that around 60\% of usability issues identified by UX-LLM are valid.

However, we noted some discrepancies in the experts evaluations.
For instance, the experts collectively identified 37 issues as usability issues, while individually they reported 27 and 31 issues.
This is highlighted by  %calculating the inter-rater reliability using the 
Cohen's Kappa measure, a widely accepted metric for assessing evaluator agreement  \cite{sun2011meta}, which was $\kappa=0.53$.
%\begin{equation*}
%\kappa = \frac{p_o - p_e}{1 - p_e} = \frac{0.76 - 0.49}{1 - 0.49} \approx 0.53
%\end{equation*}
Taking into account established benchmarks for interpreting Cohen's Kappa, such as the one proposed by Landis and Koch \cite{landis1977measurement} or Altman \cite{altman1990practical}, a $\kappa$ value of $0.53$ suggests “Moderate” agreement between the UX experts.

Looking at the differences, the main reason for disagreement seems to be the subjective nature of usability evaluation and a wide room for interpretation \cite{molich2008comparative}.
For example, the experts disagreed on issue \ref{app:C19} about the absence of a description label on the progress bar during the quiz. 
One labelled it as an issue, while the other argued that it should be intuitive to users what a progress bar in a quiz app represents.
Similarly, \ref{app:C27} mentions that after the quiz, a “more detailed feedback [despite the score] could enhance the learning experience”, which one expert viewed as a suggestion for a new feature rather than an existing usability issue.
Likewise, issue \ref{app:C32} denotes the lack of clear separation between tasks, since tasks were listed without distinct dividers or spacing in the overview of the To-Do app.
This was seen differently by experts: one classified it as a usability issue that could overwhelm users, while another considered it a minor aesthetic choice, assuming that users can navigate through lists regardless of visual grouping.
 
%{0.8\textwidth} { 
 % | >{\raggedright\arraybackslash}X 
 % | >{\centering\arraybackslash}X 
 % | >{\raggedleft\arraybackslash}X | }

%{ |p{3cm}||p{3cm}|p{3cm}|p{3cm}|  } 

Next, we calculated the precision and recall \cite{melamed2003precision} for UX-LLM.
Due to variations in the expert assessments, different metrics for each expert are calculated. 
To calculate these metrics, we need a definition for true positives ($TP$), false positives ($FP$) and false negatives ($FN$) in the context of UX-LLM. 
True positives are all usability issues generated by UX-LLM that an expert has identified as a “Usability Issue” during their evaluation.
Samples identified as “No Usability Issue” and “Incorrect/Irrelevant Statement” are defined as false positives.
The issues labeled as “Uncertain” are excluded from the calculations because the experts said that they might provide valuable insights, even if they only offer a change in perspective.
For example, issue \ref{app:C2} critiques the dark mode colour scheme, which was difficult for the experts to assess, but both agreed on the positive impact of highlighting it to developers.
False negatives are defined as all usability issues not identified by UX-LLM but found during the usability testings or expert reviews.
To avoid counting the same usability issue multiple times, we manually matched the same issues from each method, as presented in Table \ref{apptab:usabilityIssuesMatched}.
From this, the precision and recall values for Expert 1 (E1) and Expert 1 (E2) are as follows:  

\begin{align*}
Precision_{E1}  &=    &\frac{TP_{E1}}{TP_{E1} + FP_{E1}}&     &=&      &\frac{27}{27 + 17}&   &\approx 0.61 \\[8pt]
Recall_{E1}     &=    &\frac{TP_{E1}}{TP_{E1} + FN}&          &=&      &\frac{27}{27 + 51}&   &\approx 0.35 \\[8pt]
\\
Precision_{E2}  &=    &\frac{TP_{E2}}{TP_{E2} + FP_{E2}}&     &=&      &\frac{31}{31 + 16}&   &\approx 0.66 \\[8pt]
Recall_{E2}     &=    &\frac{TP_{E2}}{TP_{E2} + FN}&          &=&      &\frac{31}{31 + 51}&   &\approx 0.38 \\[8pt]
\end{align*}

To directly address research question one (\hyperlink{rq1}{RQ1}) the precision scores, which range from $0.61$ to $0.66$, are encouraging.
This suggests that the usability issues generated by UX-LLM are generally valid while maintaining false positives at a manageable level.
The recall scores, being on the lower side, ranging from $0.35$ to $0.38$, were expected given the complexity of detecting usability issues. 
This was highlighted by Molich et al.~\cite{molich2008comparative}, who demonstrated that even simple websites can contain more than a hundred usability issues.

\begin{table}
\centering
\def\arraystretch{1.1}%
\caption{Matching of issues in reference apps identified by usability testing (A), expert review (B), and UX-LLM (C).}
\label{apptab:usabilityIssuesMatched}
\begin{tabular}{  | p{1.5cm}  | p{1.2cm} | p{1.1cm} | p{1.1cm} | p{1.6cm} |   }
\hline
\textbf{View}  & \textbf{U.~Testing}  & \textbf{Expert 1} & \textbf{Expert 2}  & \textbf{UX-LLM}            \\
\hline \hline
Category & \ref{app:A1} & \ref{app:B4} & \ref{app:B4} & / \\
View & \ref{app:A2} & \ref{app:B5} & / & / \\
(Quiz App) & \ref{app:A3} & / & / & \ref{app:C6} \\
& / & / & \ref{app:B2} & \ref{app:C1}, \ref{app:C12}, \ref{app:C20}, \ref{app:C23} \\
& / & / & \ref{app:B6} & \ref{app:C4} \\
\hline
Setup View & \ref{app:A5} & \ref{app:B10}, \ref{app:B11}, \ref{app:B14} & \ref{app:B9} & \ref{app:C8}, \ref{app:C14} \\
(Quiz App) & \ref{app:A6} & / & \ref{app:B13} & / \\
& \ref{app:A7} & / & / & \ref{app:C11} \\
\hline
Quiz View & \ref{app:A9}, \ref{app:A10} & \ref{app:B21} & / & \ref{app:C15}, \ref{app:C21} \\
(Quiz App) & \ref{app:A14} & \ref{app:B20} & / & \ref{app:C22} \\
& / & \ref{app:B18} & \ref{app:B18} & \ref{app:C19} \\
\hline
Score View & \ref{app:A15}, \ref{app:A16} & \ref{app:B29} & / & \ref{app:C25} \\
(Quiz App) & / & \ref{app:B23}, \ref{app:B26} & / & \ref{app:C24} \\
\hline \hline
List View & \ref{app:A17} & / & \ref{app:B35} & / \\
(To-Do App) & \ref{app:A18} & \ref{app:B37} & \ref{app:B37} & \ref{app:C37} \\
& \ref{app:A19} & \ref{app:B36} & / & / \\
& \ref{app:A21} & / & / & \ref{app:C33} \\
& \ref{app:A22} & / & \ref{app:B43} & \ref{app:C35} \\
& / & \ref{app:B38} & / & \ref{app:C32} \\
& / & \ref{app:B39} & / & \ref{app:C36} \\ \hline
Task Detail & \ref{app:A25} & \ref{app:B57} & / & \ref{app:C44}, \ref{app:C47} \\
View & \ref{app:A26} & \ref{app:B54} & / & \ref{app:C46} \\
(To-Do App) & \ref{app:A27} & \ref{app:B58} & / & / \\
& / & \ref{app:B44} & / & \ref{app:C42} \\
& / & \ref{app:B51} & \ref{app:B51} & \ref{app:C41} \\
& / & \ref{app:B53} & / & \ref{app:C43} \\
\hline
\end{tabular}
\end{table}

\subsection{Comparison of Usability Evaluation Methods}
\label{sec:res-comparison}

To compare UX-LLM with usability testing and expert review, we matched the issues capturing the same problem possibly with a varying level of details as shown on Table \ref{apptab:usabilityIssuesMatched}.
Looking at the issues, there is a clear difference in the detail captured. 
Issues identified from usability testings generally provide a broad impression, while those from expert reviews are sharper with more precise focus.
For instance, issue \ref{app:A5} identified in a usability test,  describes that users were overwhelmed by a chaotic-looking screen. This is matched with four issues from the expert reviews: \ref{app:B9}, \ref{app:B10}, \ref{app:B11}, and \ref{app:B12} that provide more details: revealing reasons why the screen appears overwhelming such as “overload of content”, “inconsistent grid layout”, and “texts [...] inconsistent [...] use of capitalisation”. 
%Essentially, a single overarching usability issue can be deconstructed into several smaller usability issues.

\begin{figure}
  \centering
\includegraphics[width=0.9\linewidth]{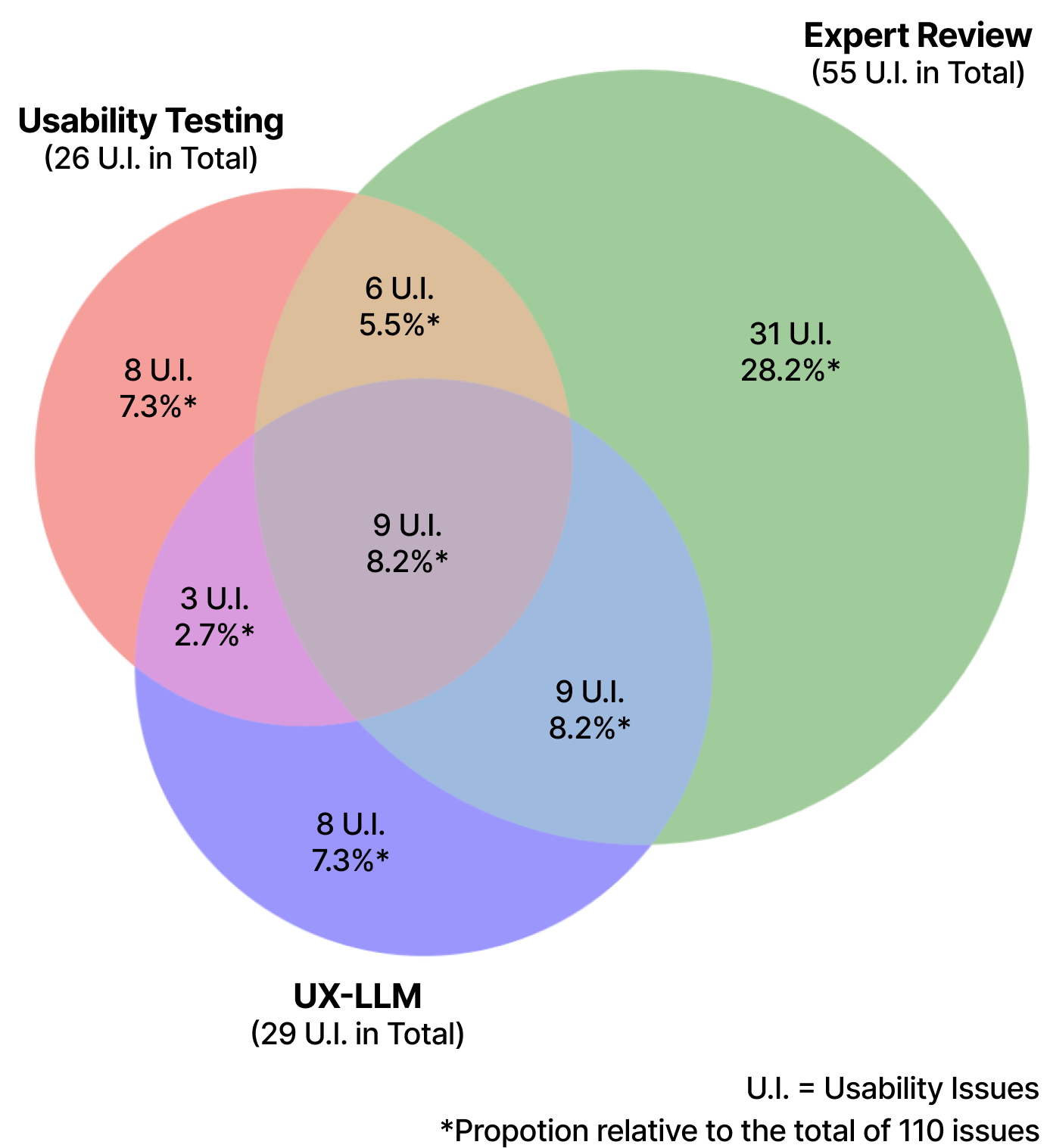}
  \caption{Venn Diagram showing the overlap of usability issues identified by usability testings, expert reviews, and UX-LLM.}
  \label{fig:Venn}
\end{figure}

The Venn diagram shown in Figure \ref{fig:Venn} is obtained by matching duplicate/similar issues.
%It is important to note that this diagram only provides a rough overview. 
%For absolute correctness, the matching would have  to be performed by UX experts, and similar issues within each method would also have required matching.
In the issue set of UX-LLM, all samples were included where \textit{at least one expert} identified them as actual usability issues. 
The diagram shows the overlap and unique contributions of  the usability testing, the expert review, and UX-LLM. 
Of the total 110 issues, the usability testings uncovered 26 issues, with 8 unique to it. 
The expert review pointed out 55, including 31 unique  issues.
UX-LLM identified 29 issues, contributing 8 unique insights. 
The diagram shows that only 9 issues were identified by all three methods. 
There are 6 issues that both the testings and the experts identified but UX-LLM did not.
The usability testing and UX-LLM together identified 3 issues that the expert review missed. 
Similarly, the expert review and UX-LLM together found 9 issues that were not detected through the usability testing. 

% If footprint should be added to updated figure:
%  \caption[Caption for LOF]{Venn Diagram showing the overlap of usability issues identified by usability testings, expert reviews, and UX-LLM.\footnotemark}
%  \label{fig:Venn}
%\end{figure}
%\footnotetext{\phantom{a}Diagram was updated because the total number of usability issues across the three methods was off by one. This correction does not affect the results.}
%We highlight several remarkable observations. First, looking at the unified overlap between the three methods, it could be assumed that those issues are particularly severe or detectable. Future studies could focus on assessing whether these issues have a higher severity or detectability than others, for example, by conducting an issue severity rating using criteria developed by Nielsen \cite{nielsen1995conduct}, and Tullis and Albert \cite{tullis2013measuring}.

It is noteworthy that the \textbf{expert review} identified a greater number of distinct issues. 
They break down overarching usability issues into several smaller usability issues, as in the example above (B9-B12).
This allows for more targeted improvements, as more focused issues make  resolution more actionable \cite{Bettenburg:FSE:2008,Martens:RE:19}.
Further, they point out minor design issues, such as a label being slightly misaligned vertically (\ref{app:B24}), or minor copy-writing inconsistencies, where the same feature is referred to with different words (\ref{app:B30}).
These issues likely went unnoticed in the usability testing as they were too minor to significantly impact the user experience for an evaluator to notice (but are certainly still relevant). %(they can still negatively impact the user experience).
%However, this does not imply that such issues should be ignored, as they can still negatively impact the user experience.

The intersection between UX-LLM and the expert review,  seems to contain issues that are difficult to extract from usability testings due to the absence of specific user groups or testing conditions.
For instance, issue \ref{app:C1} refers to a text that is hard to read for “users with visual impairments or when viewing in bright light”.
Another type of issues identified are those that could not be stimulated with the the usability testing tasks.
For example, issue \ref{app:B38} notes that “Completed tasks do not visually stand out from pending tasks”.
This might have gone unnoticed in the testings, as the maximum number of tasks displayed on the overview was four and the problem might become significant only as the list expands.

Finally, looking at the issues \textbf{only UX-LLM} has identified, it seems to spot issues other methods might miss by analysing code. 
It identifies, e.g., problems on less common user paths, such as the app  performance under slow internet conditions.
For example, issue \ref{app:C7} states: “When the app is fetching categories, there is no visual indication to the user that data is being loaded, which could lead to confusion”. 
This scenario was neither encountered in the usability testing nor the expert reviews, as they had fast internet connections.
However, it is valid for users with a limited internet connection. 
Another example is issue \ref{app:C18}, which notes: “The question text does not have a maximum line limit, which may cause issues if the question is too long, leading to the text being cut off or the layout looking cluttered”.
This issue was also not encountered because the test questions were not overly long. 
With different longer questions, this could become a usability issue.
However, UX-LLM did not identify broad context or navigation-related issues, likely since it analyses only one view at a time.
Issue \ref{app:B30}, e.g., that involves inconsistent wording of the same feature across different screens, was beyond UX-LLM capability.

%To address the second research question (\hyperlink{rq2}{RQ2}), the findings show that there are overlaps among user testing, expert review, and UX-LLM in identifying usability issues, yet each method also identified distinct issues.
%Notably, UX-LLM identifies unique issues in less common user paths, such as error paths, due to its ability to analyze the source code.
%Consequently, the results indicate that UX-LLM cannot completely replace traditional evaluation methods, but can serve as a valuable supplement.

% \begin{table*}
% \centering
% \def\arraystretch{1.3}%
% \begin{tabularx}{\textwidth}{|c|X|c|}
% \hline
% ID & Question & Ratings \\
% \hline
% \ref{app:D5} & On a scale of 1 to 4, how would you rate your initial impression of UX-LLM when it comes to improving the quality of an app's user experience? & 2, 2, 3 \\
% \hline
% \ref{app:D7} & On a scale of 1 to 4, how useful do you find these usability issues? & 4, 3, 3 \\
% \hline
% \ref{app:D10} & On a scale of 1 to 4, how easy do you think it would be to integrate UX-LLM into your existing development workflow? & 2, 3, 3 \\
% \hline
% \end{tabularx}
% \caption{Ratings From Interview Participants for Likert Scale Questions}
% \label{tab:survey_ratings}
% \end{table*}

\subsection{Perception of a Development Team (RQ3)}

%We summarise the main insights from the focus group.

\subsubsection{Current Practices of the Team}
The focus group  started with a waterfall-like approach, initially creating a high-fidelity prototype, which then served as the foundation for development.
Later, the prototype was discarded and design changes were directly implemented in the app, bypassing adjustments to the design files. 
In addition, they conducted several usability evaluations through testings and walk-through, engaging directly with representatives of the visually impaired community.

\subsubsection{Predicting Usability Issues with UX-LLM}
During the interview, we briefly demonstrated UX-LLM to the team  by using it on one of their app views. 
Their initial impression of its potential to enhance the user experience of an app was moderately positive, with an average rating of $2.3/4$. 
One participant stated: “Some issues feel a bit generic and some don't make sense, since they are addressed in previous screens”.
More specifically, UX-LLM criticised the text size for being too small, which was just personalised in the onboarding before.
The other participant said: “I appreciate the fresh perspectives it offers. Even  incorrect usability issues can be valuable as they make me reevaluate design decisions” 
(referring to an issue where UX-LLM criticise the colour scheme as not  accessible). 

We examined all issues identified by UX-LLM within the seven main app views (removing results that were clearly false positives).
The students evaluated the \textbf{usefulness} of these  issues, resulting in a positive average rating of $3.3/4$. 
They found some of the issues particularly insightful, even uncovering bugs they had not recognised before.
One participant said: “The feedback on the button bug was spot on; it's not something we  would have thought about by ourselves”.
The other acknowledged that the issues are properly phrased so that they can “directly be imported as to-dos on our task board”.
Also, “On some screens we assumed something is not ideal, but we did not know what the problem was, these issues are very helpful”.
Overall, they felt that pre-filtered results significantly improved the usefulness and appeal of UX-LLM.

\subsubsection{Integrating UX-LLM into Development Workflows}
%We discussed the integration of UX-LLM into the team workflows.
Regarding the ease of integrating UX-LLM, the team gave it a mildly positive average rating of $2.6/4$.  
A developer mentioned that it would be a further burden for him to fill out UX-LLM required fields.  
He stated: “I'm a laid-back person, so it would annoy me to have to use another application beside my IDE”.
In addition, he said: “I'm not sure if I'd regularly go through these $\sim$10 points”, 
indicating that it could be tedious to go over the issues for every app view over the course of development.

The group considered having a team member consistently review the app views with UX-LLM and share the feedback with the rest. 
One of them said, “Personally, I wouldn't mind using the tool, as I, being the project manager, could regularly use UX-LLM and incorporate its findings into our task board”.
Although doable with their team size, they recognised the challenge of adopting another task given their current time pressure.
The UX designer stated, “It's great to see an overview of what's available; you can quickly eliminate unnecessary issues and reflect on them.
In the end, it saves a lot of time as it is easier than conducting usability evaluations ourselves”, 
indicating her excitement about the tool and that she feels like she can easily filter out the false positives.

\subsubsection{Additional concerns and improvement suggestions} For applying UX-LLM in daily  work, the group proposed embedding it as a plugin in the IDE or in a continuous integration (CI) pipeline to allow for automatic operation in the background.  
They emphasised that an IDE plugin “just makes sense, as the Xcode IDE\footnote{\url{https://developer.apple.com/xcode/}} 
already has the code and view's preview side by side”.
In addition, the team expressed the desire to not only identify issues but also  provide solutions, including alternative designs of their input image.
The designer said: “When it criticised the accessibility of the colours, it would be nice if it could also show what colours to use instead”.
Interestingly, they also stated a limitation that could result from the policy of the LLM used.
A discussed example is when creating an app about wine, as some LLMs might not be allowed to answer queries about alcohol.
Furthermore, the presentation of usability issues was relevant to the team, as they found it demotivating to receive a lengthy list pointing out their product deficiencies.  
Also, the developer was sceptical about UX-LLM capability to identify usability issues in features reliant on hardware interactions involving user input or external factors, such as camera functions and voice input.
Lastly, they emphasised the importance of a holistic analysis  to detect broader inconsistencies and navigation issues. 
This approach would also reduce predicted usability issues, which are already addressed in other sections of the app.

Overall, the team perspective on UX-LLM was positive, ending the session saying: “It has identified issues that we overlooked, and not just a few”.
The tool was credited with uncovering valuable issues that had previously gone unnoticed, besides conducting usability evaluation methods.

\section{Related Work}
\label{sec:relatedWork}
There is a huge body of knowledge around usability \cite{frokjaer2000measuring, iso9241_1998, iso9241_2018} and usability engineering \cite{nielsen1994usability, arnesson2012usability, harrison2013usability}, including a recent comprehensive review of usability for mobile apps \cite{weichbroth2020usability}.
The closest areas to our work are
software analysis for usability, user data and feedback analytics, and GenAI for GUI development.

\subsubsection{Code Analysis for Usability}

Research has proposed static and dynamic code analysis to identify usability issues. 
%\subsection{Code Analysis}
For example, Mathur et al.~\cite{mathur2018usability} proposed an analysis framework to check the source code of Android apps against (pre-configured) usability guidelines and validation cases (e.g.~all password fields should provide an option to reveal the clear text).
Similarly, UX-LLM only uses development artefacts, notably source code, but focuses on iOS apps and does not require pre-configuration. 
In a study with 16 mobile app companies, Mathur et al.~found that 
developers often skip usability evaluation due to needed effort, lack of resources, and know-how \cite{mathur2018usability}. 
This is a main motivation for our AI-assisted usability evaluation.  
The authors also found their framework to be effective and helpful but developers were sceptical about the extra work to build own validation cases. 
%We also made similar observations in our focus group.  
Similar methods \cite{Abuaddous:IMT:2022, Coppola:JSS:2021} are capable of detecting pre-defined issue (categories), which can be beneficial, e.g., when evaluating accessibility requirements \cite{Bajammal:ICSE:2021, Alshayban:FSE:22, Silva:BSST:2020}.
However, they require initial setup and maintenance as development libraries, guidelines, and usage scenarios evolve over time \cite{softwarechange}.
%In contrast, UX-LLM and the approach of Jongwook et al. operate more exploratory and do not require prior setup, making them easier to use and potentially allowing for a broader spectrum of usability issues to be detected.
As UX-LLM the main advantage of  code analysis approaches is their applicability during the development phase to predict and prevent usability issues, before such issues impact real users.

\subsubsection{Usage Data Analytics}
App usage logs deliver great value for developers as they enable analysing the user behaviour \cite{Maalej:Software:2016, Gomez:Software:2017, Roehm:ICPC:13}.  
For example, they can identify which app features are used most and focus usability evaluation on those areas \cite{Stanik:RE:2020, harty2021logging}.
Mobile tracking \cite{iacob2013retrieving, harty2021logging} aka analytics tools like \textit{Firebase}
\footnote{\url{https://firebase.google.com}}
or \textit{TelemetryDeck}
\footnote{\url{https://telemetrydeck.com}}
enable exploring runtime, engagement, and navigation data, or tracking actions such as button presses. 
However, such tracking remains exploratory and raises privacy concerns depending on its implementation \cite{tang2019demystifying, VanDerSype:RELAW:14}.
 
To detect issues, Park et al.~\cite{park2018automatic} generated a representative “real user activity model” from execution traces and an “expected activity model” from developers' execution of intended usage scenarios.  
The authors matched both  models to reveal discrepancies between intended and actual usage,  uncovering four types of usability issues: “unexpected action sequence”, “unexpected gesture”, “repeated gesture”, and “exceeded elapsed time”. 
%Park et al.~evaluated their approach with five open-source Android apps and demonstrated an average precision rate of 70\% in identifying usability issues.
Similarly, Jeong et al.~\cite{jeong2020detecting} presented a graph-based approach to identify dissimilarities in user behaviour, assuming that when users face usability issues, they show different behavioural patterns.  
Evaluating the approach on issues identified from usability testings of two Android apps, the authors found a correlation between areas of high behavioural dissimilarity and identified usability issues. 
%The also found that issues with a high severity level had low similarity scores, yet low similarity does not necessarily indicate high severity \cite{nielsen1995conduct, tullis2013measuring}. 
%\subsection{Comparison to UX-LLM}
Unlike UX-LLM, usage data analytics often does not require access to the source code, but actual users executing the app. 
This provides a strong evidence about the user issues but prevents the applicability to earlier development stages, e.g.~before releases.
Moreover, tracking user activity could raise privacy concerns and only some user segment might participate \cite{VanDerSype:RELAW:14}. 

%%%%%
\subsubsection{User Feedback Analysis}
Research has shown that, for many developers, App Store reviews represent a primary source of user feedback \cite{Hassan:EMSE:2018, pagano2013user}. 
%The user reviews address issues like bug reports and feature requests.
These reviews enable gathering issues (including on usability) and help developers improve their apps, enabling user-driven quality assessment  \cite{Maalej:Software:2016}.
Users also report issues in other feedback channels too as social media \cite{Martens:RE:19} and product forums \cite{Tizard:RE:2019}.  
%Nayebi and Cho  \cite{nayebi2018app}, e.g., revealed that more than 12\% of bug reports were discovered on Twitter%\footnote{\url{https://x.com}},  now called “X”.
Thus, monitoring and processing feedback looking for usability related reports is an active research field that has received a strong boost with the availability of recent LLMs \cite{Maalej:NLP4RE:2025, Wei:ICTAI:2023}. 
While certainly an important complementary source for monitoring usability issues,  actual user priorities, and novel ideas, user feedback analytics also requires releasing the app and having a significant user base---whereas UX-LLM can be used during the development as it simulates common (user) knowledge coded in foundation models.  
UX-LLM is also beneficial for less popular apps, which do not have enough users for a meaningful analysis \cite{pagano2013user}.
%UX-LLM specifically identifies usability issues at the view level, providing a rationale.
Moreover, feedback analytics requires manual tuning and interpretation to extract more valuable insights \cite{zhi2019exploratory, fu2016tuning, Maalej:NLP4RE:2025}. 
Feedback can vary greatly in quality \cite{Maalej:NLP4RE:2025} with up to 75\% of app store reviews consist of praise rather than constructive feedback \cite{pagano2013user, Martens:RE:19}.  
%Therefore, while analytics and user reviews can offer a wide range of insights, including usability issues, they are not exclusively focused on these.   Ultimately, UX-LLM serves as a supplementary tool to proactively identify and address usability concerns during development.
% By combining UX-LLM with other methods, developers can address usability issues pre-release and also gather valuable data from actual users post-release, offering insights into user behaviour, wishes and bugs.

\subsubsection{Generative AI for GUI Engineering}

% OpenAI's Codex is a highly powerful LLM for solving code-related tasks \cite{DBLPabs-2107-03374, nguyen2022empirical, lemieux2023codamosa}.
% It is a closed-source GPT-based model, which was fine-tuned on 54 million public GitHub repositories \cite{brown2020language}.
% Thus, Codex powers the coding assistant “GitHub Copilot”\footnote{\url{https://github.com/features/copilot}}, which generates code directly in the IDE, based on  provided context \cite{tambon2024bugs}.
% For example, it can write functions from the signature and explain its input and desired output.
% Recently added “Copilot Chat”\footnote{\url{https://docs.github.com/copilot/github-copilot-chat}} provides answers to coding-related questions, helping to generate unit test cases, explaining code, suggesting improvements, and proposing code fixes \cite{GitHub2024CopilotChat}.
Recent advance in AI has also benefited usability research and UI design in general. 
Beltramelli \cite{beltramelli2018pix2code} presented “pix2code”, which generates code from a GUI image input. 
This approach uses Convolutional Neural Networks to identify distinct elements, positions, and poses in GUI images.
It then employs Recurrent Neural Networks and a decoder to generate executable code.
The rapid raise of Generative AI  brought significant improvements in how models solve the “Design2Code” task.
Si et al.~\cite{si2024design2code}.
conducted a benchmarking with  state-of-the-art models and showed that GPT-4 Turbo with Vision performs best on this task, compared to other models like Gemini Pro.
Their study annotators claimed that GPT-4 generated websites which can replace the original reference in terms of visual appearance and content in 49\% of cases.
In 64\% of cases, the annotators stated that GPT-4 even generated better websites than the original reference. 

%\subsection{GUI Generation}
Other recent approaches has suggested to generate GUI images. 
Lee et al.~\cite{lee2020neural} proposed a hybrid neural network to generate design layouts that meet user-specified constraints (e.g.,~a design with one headline and two images).  
In their experiments, they demonstrated that the generated layouts are visually similar to real design layouts.
Wei et al.~\cite{wei2023boosting} took  a step further to generate the designs using UI-Diffuser and LayoutDM \cite{chai2023layoutdm}, a Transformer-based model for layouts generation.
Users provide a simple prompt of what the GUI should represent, e.g., a music player.
Their preliminary results indicate that this could be a cost-effective approach for creating mobile GUI designs. 
However, the GUIs are far from being directly directly reusable, need improvements, and usability evaluation.
Uizard\footnote{\url{https://uizard.io}} 
is a similar recent commercial product, which generates mobile GUI designs with a simple prompt.
Mattisson et al.~\cite{mattisson2022vad} demonstrated that Uizard can significantly enhance designers efficiency, boost creativity in idea generation, and improve communication with stakeholders.
Particularly the editable designs with nested components allow for continuous customisation and adjustments, e.g. after usability evaluation. 
In another recent study, Wei et al.~\cite{Wei:TOSEM:24} tuned a Vision-Language Model to retrieve relevant GUI designs from a large constructed dataset. 
This approach seems to be more effective yielding more relevant and higher quality screens---highlighting the importance of the human-in-the-loop for GUI recommendation.

The works are are perhaps closest to ours are of Kuang et al. \cite{Kuang:CHI:2024} and of Kocaballi \cite{Kocaballi:2023}.
To explore how UX evaluators use AI assistant Kuang et al.\cite{Kuang:CHI:2024} conducted a study with 20 participants using a simulated AI assistants. 
They found that participants asked the bot for five categories of information such as user actions and mental model and observed other trends.  
The authors finally derive design considerations for future conversational AI assistants for UX evaluation. 
In hypothetical design project, Kocaballi used ChatGPT to generate personas, simulate interviews with fictional users, and create new design ideas. 
The work highlights  drawbacks such as forgotten information, partial responses, and a lack of output diversity, suggesting the importance of human oversight in the design process.
To the best of our knowledge, this work is the first to use Generative AI for predicting actual usability issues in apps while comparing the performance with conventional usability testing and evaluation by humans. 

%we  argue that the advantages of using LLMs lie in their broad applicability and their ability to provide more than just numerical scores, as they offer rationales for the usability issues they identify, enhancing the interpretability and actionability of their findings.
%Furthermore, with the continued rapid advancements in the field of LLMs, the performance of UX-LLM could be easily improved by simply exchanging the underlying models while still maintaining the same architecture \cite{zhao2023survey}.

\section{Threats to Validity and Limitations}\label{sec:threats}
%In the following sections, the internal, external, and construct validity of the studies that were used to answer the research questions are highlighted.

\subsubsection{Internal and External Validity}
For the usability testing, we aimed to recruit diverse  participants, with ages ranging from 24 to 60 years and various occupations.
Although this represents a broad spectrum, some participants have a technical affinity, which could potentially limit the range of usability issues identified.
Furthermore, only three  participants were female and three had only limited iOS experience, which could potentially obscure the results, as unfamiliarity with platform features might be mistakenly perceived as usability issues \cite{pong2019awareness}.
In addition, no participant had impairments, potentially missing some issues. 
Replicating exact real-world conditions is a perpetual challenge.
Although usability tests (also ours) are designed to match real-world conditions (than laboratory tests), they are never entirely the same \cite{kallio2005usability}.
Some usability issues might have been overlooked due to  artificial usage environment or participants motivation to use the apps with pre-defined tasks.

Another potential threat to internal validity is that only two UX professionals were recruited for the expert review, %and particularly only one development team for the focus group.
Although the evaluations were thoroughly conducted as the result details show, involving additional experts might lead to more diverse perspectives on UX-LLM and its results. 

%\subsubsection{External Validity}
The selection of two reference apps restricts a broader reliability of our findings.
The chosen Quiz and To-Do app are simpler compared to the wide range of  more complex available apps.
As we primarily aimed to support smaller app teams, it made more sense to focus the evaluation on simpler apps. 
Evaluating UX-LLM for niche domains like ``Ireland - COVID Tracker'' would rather limit the results.
The two apps used in our study provided the best compromise between: 
\begin{itemize}
    \item Broadly used and maintained app.
    \item Clean SwiftUI Model-View-ViewModel Architecture to inject into UX-LLM.
    \item General apps, not too niche domain (also to enable the usability testing).
\end{itemize}.
Nevertheless, to increase the generalisability, it is important to replicate our study with apps from various categories, embodying varying levels of complexity and incorporating different types of user interactions \cite{Roehm:ICPC:13, Maalej:RSSEBook:2014}.
The amount and quality of code entered into UX-LLM can also have a major impact on its performance.

%Therefore, it remains uncertain to what extent UX-LLM can accurately predict usability issues across a broader spectrum of apps.

We chose expert reviews and usability testings as comparative usability evaluation methods. 
While these methods are well-established \cite{paz2016systematic}, for a comprehensive analysis, UX-LLM should also be compared with additional usability evaluation methods such as questionnaires or feedback, usage data \cite{Roehm:ICPC:13}, and code analysis techniques discussed in Section \ref{sec:relatedWork}. 

In addition, the focus group  involved a small student development team working on a capstone university project app. 
The insights gained do not fully represent the experiences of more diverse professional development teams working on a commercial app, affecting the perceived utility and integration potential of UX-LLM in various development contexts. 
For more in-depth results, it would be beneficial to  have developers use the tool over time to gather more evaluation experiences.

\subsubsection{Construct Validity} 
Unlike the usability testing and the expert assessment which followed strict protocols, during the expert reviews we refrained from instructing the experts to follow a certain process. 
A more structured review would likely increase the replicability and agreement rate. 
In fact, we considered using the more structured approach called ``Heuristic Expert Reviews'' where evaluators are aided by a checklist.
However, we decided for the general review form: mainly to not interfere or direct the experts to a certain direction \cite{Harley:2018}. 
Moreover, heuristic evaluations tend to miss context-specific issues \cite{Harley:2018}. 

During the expert assessment, the survey had four options to categorise UX-LLM's usability issues: “Usability Issue”, “No Usability Issue”, “Uncertain”, and “Incorrect/Irrelevant Statement”. 
The appropriateness of these categories could also be questioned, considering that they might not capture all possible reactions.
To mitigate this potential threat, we took detailed notes of all expert comments. 

Furthermore, during performance evaluation, the recall was calculated by defining false negatives as all usability issues not identified by UX-LLM but found during usability testings and expert reviews. 
There remains ambiguity about whether all usability issues have been uncovered, with a strong likelihood that some issues have gone unnoticed.
This was highlighted by Molich et al. \cite{molich2008comparative}, who demonstrated that even simple websites can contain more than a hundred usability issues.
Similar to Park et al.~multiple studies skipped the recall calculations entirely, stating that, “it is nearly impossible to define a totally complete set of [usability issues].'' \cite{park2018automatic} 
Accordingly, we combined issues from both expert evaluation and the testings suggest using precision as main metric to demonstrate the method accuracy and recall rather as indicative.

To compare and match the overlap of usability issues found by the three methods, we constructed Table \ref{apptab:usabilityIssuesMatched}.
Although we took special care and applied a strict guideline that only the same or sub/super issues are matched, other UX experts might come to slightly different matching  \cite{molich2008comparative, jacobsen1998evaluator}, which we encourage assessing by share our evaluation data.  
Only usability issues between the methods were matched. Matching issues within each method could also lead to additional insights.

Finally, for the studies on UX-LLM only OpenAI's GPT-4 Turbo with Vision was used.
Using different LLMs and different prompts could significantly change the performance. 
Moreover, during all studies, we ignored the severity of issues, 
Comparing the issue severity would likely lead to additional insights but requires additional studies, due to its subjective variability  \cite{jacobsen1998evaluator}.
We also did not examine the determinism of UX-LLM results. Generating different outputs from the same input could impact performance and usefulness \cite{tian2023chatgpt}.

% \section{Conclusion}
% \label{sec:conclusion}
\section{Discussion and Conclusion}
\label{sec:discussion}\label{sec:conclusion}

%

%Integration into the IDE as a debugger . 

Our results show that---with fairly low effort and widely-available models---GenAI can predict in source code valid usability issues, that can easily be reviewed and fixed before releasing the app, avoiding to dissatisfy users and compromise their experience. 
This saves time and resources particularly for smaller app teams and freelancers, who tend to ignore usability evaluation \cite{mathur2018usability}. 
However, our study also shows that relevant issues particularly identified by usability experts were missed by UX-LLM. 
This suggests that traditional usability evaluation can certainly be boosted with GenAI but not completely  replaced, still requiring a \textbf{UX expert in the loop} \cite{Wei:TOSEM:24}.  

Usability evaluation is a rather continuous, iterative, subjective process  \cite{nielsen1994usability,nielsen2012usability}. 
It is thus unlikely that one method or one person can identify ``all'' usability issues \cite{molich2008comparative}. 
This should be kept in mind when interpreting the rather limited recall values of UX-LLM, calculated conservatively. 
Relaxing the interpretation, e.g., including issues rated ``undecided'' or confirmed by one expert would lead to higher performance values.   
Our qualitative analysis, e.g., from the focus group, also shows that ``debatable'' issues or ``rather feature ideas'' can also be insightful for developers to re-evaluate their designs. 

We think that \textbf{method triangulation} or hybrid approaches are particularly appealing to gather various perspectives and increase evaluation effectiveness and efficiency.  
UX-LLM seems to outperform when access to certain users is difficult (e.g.~impaired users) or rare scenarios and tasks are not specified (Section \ref{sec:res-comparison}).
Experts are particularly able to reason about the broader pictures, e.g., issues across multiple views or inconsistent terminology, which is a limitation of UX-LLM. 
Automated analysis of user feedback, usage data, and code discussed in Section \ref{sec:relatedWork} represent additional complementary techniques. 
However, a central question remains unanswered: how and when to effectively combine the various usability evaluation techniques. 
Our comparative study and the focus group  contributed only preliminary insights.  
This need to be re-evaluated at a broader scale and including multiple apps, domains, programming frameworks, and numerous issues. 

At the time of our research, there was \textbf{no public datasets} available with source code and actual validated usability issues from various evaluation methods. 
This was a major rationale to focus our work on an in-depth study creating such data for 3 apps, rather than a benchmarking different Foundation Models and prompt engineering techniques. 
We hope that by replicating our study and augmenting our dataset with additional apps and issues, studies on model-tuning and advance prompt engineering (including, e.g. more / custom issue examples or techniques as Chain of Thoughts) will be become possible. 
This will likely lead to more precise GenAI-assisted usability evaluation tools  
and will enable meta-studies of hybrid approaches. 
We think that the research community should focus on this in near future.
Finally, integrating usability issue prediction in the IDE (as a ``usability debugger'') as well as suggesting potential fixes are promising directions for a higher acceptance of the tool in daily development work.

% Although, they criticised that it needs filtering as some issues are generic, wrong, or irrelevant, partly because UX-LLM does not see the other parts of the app besides the provided view.
% They appreciated the fresh perspective provided by UX-LLM and noted that it initiates valuable re-evaluation, especially when being uncertain about design decisions.
% However, they also expressed concerns about the practicality of integrating UX-LLM into existing workflows, especially due to the additional overhead of interacting with another tool.
% Finally, they suggested improvements for future iterations, where UX-LLM could be better integrated by automating certain steps.

%To address research question three (\hyperlink{rq3}{RQ3}), initial contact with emerging developers shows that UX-LLM is perceived positively, indicating that it has the potential to support development by identifying usability issues. 
%Additionally, the interview highlighted areas for improvement, especially when it comes to integrating UX-LLM into development workflows.
%For significant results, additional interviews with development teams of varying sizes and levels of professionalism are necessary.
%Nevertheless, the overall direction appears promising, as indicated by the feedback.

\section*{Acknowledgement}
This work was partly supported by MaibornWolff GmbH. 
We thank all participants in the usability testing, the UX experts, as well as the participants of the focus group for their time and feedback.

% Bibliography

\IEEEtriggeratref{68}
%\afterpage{\blankpage} 
\bibliographystyle{IEEEtran}
\bibliography{main}
%\afterpage{\blankpage} 

%\input{Resources/Pages/disclaimer.tex}

\newpage

\section*{Usability Issues from Usability Testing} \label{appendix:usabilityTesting}

\begin{enumerate}[label={}]

\section*{Quiz App}
    \item \textbf{Category View}
    \begin{enumerate}[label=A\arabic*, ref=A\arabic*]
        \item Users did not immediately recognize what the view is about, requiring them to read the categories to understand it. \label{app:A1}
        \item Users were confused by the sorting of the categories, as they could not identify a clear sorting rule, such as alphabetical order. \label{app:A2}
        \item The light/dark mode toggle was misleading; users expected the icon to change from a sun to a moon when toggled. \label{app:A3}
        \item Difficulty was encountered in finding the “Geography” category due to the lengthy list without a search function. \label{app:A4}
    \end{enumerate}
    \item \textbf{Setup View}
    \begin{enumerate}[resume, label=A\arabic*, ref=A\arabic*]
        \item The design was perceived as chaotic. \label{app:A5}
        \item Uncertainty regarding “Any” as a difficulty level and its emoji. \label{app:A6}
        \item Delay after pressing the “Start” button led to multiple presses. \label{app:A7}
        \item The back button was misleadingly labeled as “Back” instead of something more descriptive like “Change Topic”, causing navigation confusion for some users. \label{app:A8}
    \end{enumerate}
    \item \textbf{Quiz View}
    \begin{enumerate}[resume, label=A\arabic*, ref=A\arabic*]
        \item Some users were unsure which answer they had logged in and whether it was correct. This is due to the sudden change of the screen, accompanied by green, red, and yellow indicators, along with multiple symbols appearing. \label{app:A9}
        \item The cross and checkmark symbol caused confusion among users. \label{app:A10}
        \item Text changed from one line to two lines after displaying the cross / checkmark symbol. \label{app:A11}
        \item When “Any” difficulty was chosen, users wanted to know the difficulty level of questions shown to them. \label{app:A12}
        \item Current Points were not visible to the user, leading to uncertainty about their score. \label{app:A13}
        \item Users were unable to correct their answers when they accidentally pressed the wrong button. \label{app:A14}
    \end{enumerate}
    \item \textbf{Score View}
    \begin{enumerate}[resume, label=A\arabic*, ref=A\arabic*]
        \item Users who pressed “Try again” to answer different questions on the same topic were surprised to receive the same questions again. \label{app:A15}
        \item Pressing “Back” led users to the setup screen instead of the categories screen, which was not as expected. \label{app:A16}
    \end{enumerate}

\section*{To-Do App}
    \item \textbf{List View}
    \begin{enumerate}[resume, label=A\arabic*, ref=A\arabic*]
        \item When trying to tap the “Add To-Do Task” button was often missed due to its small size and placement at the top of the screen. \label{app:A17}
        \item Task color coding was unclear, particularly the use of blue instead of orange, to follow the expected traffic light colors of red, orange, green for high, medium, and low priorities. \label{app:A18}
        \item Task sorting by priority was not clear until tasks of all priority levels were displayed. \label{app:A19}
        \item Accidentally marking tasks as completed instead of opening the edit view. \label{app:A20}
        \item Some users cleared tasks individually by swiping, overlooking the option to use the trash bin for clearing all tasks at once. \label{app:A21}
        \item Some users were unable to find the swipe-to-delete feature for individual tasks, instead trying a long press or searching for an edit mode button. \label{app:A22}
        \item Uncertainty about the disappearance of completed tasks (e.g., if they disappear automatically after a certain time). \label{app:A23}
    \end{enumerate}
    \item \textbf{Task Detail View}
    \begin{enumerate}[resume, label=A\arabic*, ref=A\arabic*]
        \item Tendency to create a single task for all activities (e.g., “Daily Tasks”) due to lack of subtasks or examples, leading to overloaded task descriptions. \label{app:A24}
        \item It was not clear to all users about the optional nature of the task description field. \label{app:A25}
        \item Lack of clarity on why reminders were marked as invalid when accidentally set to past dates. \label{app:A26}
        \item Desire for a delete option within the task editing interface. \label{app:A27}
     \end{enumerate}
\end{enumerate}
\section*{Usability Issues from Expert Review} \label{appendix:expertReview}

\begin{enumerate}[label={}]
    \section*{Quiz App}
    
    \item \textbf{Category View}
    \begin{enumerate}[label=B\arabic*, ref=B\arabic*]
        \item The app defaults to light mode, irrespective of the device's preferences. \label{app:B1}
        \item The yellow and white color scheme of the app make text hard to read. \label{app:B2}
        \item The spacing between the list and heading is too narrow to effectively create a hierarchy. \label{app:B3}
        \item The title “Trivia Game” does not clearly indicate the screens purpose, forcing users to explore the content for clarity. \label{app:B4}
        \item The order of the categories appears arbitrary. \label{app:B5}
        \item Additional explanations for some categories might be beneficial, such as clarifying that “General Knowledge” includes a mix of questions from all categories. \label{app:B6}
    \end{enumerate}
    \item \textbf{Setup View}
    \begin{enumerate}[resume, label=B\arabic*, ref=B\arabic*]
        \item The previously selected category and its corresponding title on the setup screen sometimes do not match (e.g., “Video Games” vs. “Entertainment: Video Games”). \label{app:B7}
        \item The back button's blue accent clashes with the app's yellow accent color. \label{app:B8}
        \item There is an overload of content on a single view. \label{app:B9}
        \item Inconsistent grid layouts cause visual confusion. \label{app:B10}
        \item The lack of visual guides makes the screen appear disorganized, with buttons missing defined tap areas. \label{app:B11}
        \item Texts are inconsistent in their use of capitalization. \label{app:B12}
        \item It is unclear whether the “Any” option randomizes difficulty per question or for the entire quiz. \label{app:B13}
        \item The “Multiple Choice” text wrapping is visually unappealing. \label{app:B14}
        \item The start button is slimmer and has a shadow, leading to a lack of consistency in button design. \label{app:B15}
        \item Saving the last setup configuration could simplify repeated use. \label{app:B16}
    \end{enumerate}
    \item \textbf{Quiz View}
    \begin{enumerate}[resume, label=B\arabic*, ref=B\arabic*]
        \item There is no option to exit a game in progress. \label{app:B17}
        \item The progress label (“1/10”) could be confusing as there is no textual description of what the numbers represent. \label{app:B18}
        \item The design of the answer buttons is not consistent with the rest of the button designs. \label{app:B19}
        \item Once an answer is selected, it is instantly submitted, with no option to revise it. \label{app:B20}
        \item The presentation logic of the symbols for correct and incorrect answer (checkmark and cross) can be confusing. \label{app:B21}
        \item When submitting an incorrect answer, the black text on the red background of the answer button creates a contrast that is difficult to read. \label{app:B22}
    \end{enumerate}
    \item \textbf{Score View}
    \begin{enumerate}[resume, label=B\arabic*, ref=B\arabic*]
        \item The visual hierarchy is unclear, making it hard to distinguish the importance of different elements. \label{app:B23}
        \item The alignment of the “10/10” label to the progress indicator is not centered. \label{app:B24}
        \item The “10/10” label and progress indicator are not relevant on the result screen. \label{app:B25}
        \item Text elements are overly large and clustered at the top, lacking a balanced distribution. \label{app:B26}
        \item The positioning of buttons towards the top of the screen makes them less accessible. \label{app:B27}
        \item The design does not effectively utilize secondary buttons to indicate less important actions. \label{app:B28}
        \item The “Try Again” feature restarts with the game with the same questions, which could be clearer in its description. \label{app:B29}
    \end{enumerate}
    
    \newpage
    \section*{To-Do App}
    
    \item \textbf{List View}
    \begin{enumerate}[resume, label=B\arabic*, ref=B\arabic*]
        \item The app's interchangeable use of “Tasks”, “Things”, and “Items” creates terminology inconsistency. \label{app:B30}
        \item The wording of the headline “Things to Do” can confuse, especially in the empty state. \label{app:B31}
        \item During the empty state, showing a disabled trashcan is confusing. \label{app:B32}
        \item There is a lack of clear visual hierarchy due to the font sizes in the empty state. \label{app:B33}
        \item Inconsistent icon sizes detract from the app's visual coherence. \label{app:B34}
        \item Despite being a key function, the add task button is small and positioned in the toolbar, away from easy thumb access. \label{app:B35}
        \item The task's sorting is unclear and unexplained. \label{app:B36}
        \item The color coding for task priority does not follow the expected traffic light scheme. \label{app:B37}
        \item Completed tasks do not visually stand out enough from pending tasks. \label{app:B38}
        \item Once tasks are completed, their details become inaccessible. \label{app:B39}
        \item The search feature is hidden behind a swipe-down gesture. \label{app:B40}
        \item The search feature's empty state has no informative content. \label{app:B41}
        \item The absence of a confirmation step for swipe-to-delete actions could result in accidental task removal. \label{app:B42}
        \item An expected “select to delete” functionality is missing. \label{app:B43}
    \end{enumerate}
    \item \textbf{Task Detail View}
    \begin{enumerate}[resume, label=B\arabic*, ref=B\arabic*]
        \item The view's title “Edit Task” when adding a task is misleading. \label{app:B44}
        \item The back button's label “Things to Do” is unnecessarily long and not straightforward. \label{app:B45}
        \item The word “Title” is redundantly used both as the label and the placeholder for the title field. \label{app:B46}
        \item Unlike the other fields, the description text field contains no placeholder. \label{app:B47}
        \item The description field's dynamic resizing is unconventional and might not guide user input as effectively as a fixed size would. \label{app:B48}
        \item The ability to dismiss the keyboard by tapping on the background is missing. \label{app:B49}
        \item The arrow icons in the priority field do not clearly indicate a dropdown menu. \label{app:B50}
        \item The use of gray color for the text in the priority field misleadingly suggests that it is disabled or a placeholder. \label{app:B51}
        \item The word “Priority” is redundantly used both as the label and the placeholder for the priority field. \label{app:B52}
        \item The “Remove Reminder” button is too prominent and could be mistakenly perceived as a save button. \label{app:B53}
        \item The alert message for reminders set in the past does not clearly explain the problem. \label{app:B54}
        \item Alerts for invalid reminder dates should be given immediately when an invalid date is selected, not upon saving. \label{app:B55}
        \item The action for saving should be labeled as “Save” instead of “Done”. \label{app:B56}
        \item It is unclear that the title field is required for task creation and that the “Done” toolbar button is grayed out until it is filled in. \label{app:B57}
        \item A delete option within the task detail view is missing. \label{app:B58}
    \end{enumerate}
\end{enumerate}
\section*{Usability Issues from UX-LLM} \label{appendix:usabilityIssuesUXLLM}

\begin{enumerate}[label={}]
\section*{Quiz App}
    \item \textbf{Category View}
        \begin{enumerate}[label=C\arabic*, ref=C\arabic*]
            \item Insufficient contrast between text and background color: The yellow background with white text on the category buttons may not provide enough contrast for users with visual impairments or when viewing in bright light conditions. \label{app:C1}
            \item Inconsistent use of dark mode: The toggle for light and dark mode changes the overall UI theme but retains a yellow background for category buttons, which could lead to a disjointed user experience in dark mode. \label{app:C2}
            \item No visual feedback on button press: The category buttons do not appear to have any visual feedback when tapped, which could leave users uncertain whether their input has been registered.  \label{app:C3}
            \item Lack of category description: The buttons only display the category name without any description, which might not be enough information for users to decide which category to choose, especially if the category names are not self-explanatory. \label{app:C4}
            \item Fixed button height: The category buttons have a fixed height, which might not accommodate category names of varying lengths or localization into other languages where the text might be longer.  \label{app:C5}
            \item Potential for hidden navigation bar elements: The sun icon for toggling light/dark mode might be hard to find for users, as it blends with the navigation bar elements and does not communicate its function clearly without text or a more descriptive icon. \label{app:C6}
            \item No indication of loading state: When the app is fetching categories, there is no visual indication to the user that data is being loaded, which could lead to confusion if the process takes longer than expected. \label{app:C7}
        \end{enumerate}
        \newpage
    \item \textbf{Setup View}
        \begin{enumerate}[resume, label=C\arabic*, ref=C\arabic*]
            \item Inconsistent option button layout: The “True/False” option button for the “Question Type” is not aligned with the other buttons, which may confuse users about whether it is selectable or part of another group. \label{app:C8}
            \item Lack of visual feedback on user interactions: There is no indication that a user's selection has been registered aside from the color change, which might not be enough for users with visual impairments or those not perceptive to subtle changes. \label{app:C9}
            \item Unclear error messaging position: The error message appears at the bottom of the view, which might not be noticed by users if they do not scroll down or if there is no indication that an error has occurred. \label{app:C10}
            \item Ambiguous loading state: The loading indicator is placed on top of the “Start!” button when fetching questions, which might make users think that the button is disabled or that the app has frozen, rather than understanding that a process is ongoing. \label{app:C11}
            \item The “Start!” button's text color is similar to the background color, which may reduce contrast and readability, potentially making it difficult for users to identify the call-to-action button. \label{app:C12}
            \item The app view lacks a clear indication of the current step or progress within the quiz configuration process, which might leave users uncertain about how far along they are in the setup. \label{app:C13}
            \item No clear visual hierarchy between different sections (Difficulty, Question Type, Number of Questions), which can make the view appear as one large block of content, potentially overwhelming users. \label{app:C14}
        \end{enumerate}
    \item \textbf{Quiz View}
        \begin{enumerate}[resume, label=C\arabic*, ref=C\arabic*]
            \item Inconsistent feedback on selection: The visual feedback for correct and incorrect answers is inconsistent. The correct answer is only indicated by the color green with a checkmark, while the incorrect answer is indicated by the color red with an “X” and also shifts to the left, which may be confusing for the user. \label{app:C15}
            \item Lack of clear navigation cues: There is no clear indication of how to proceed to the next question after a selection is made, which could lead to user confusion. The “Next” button appears only after a selection, and it may not be immediately obvious to the user. \label{app:C16}
            \item Inadequate touch targets: The “Next” button may have an inadequate touch target size, which could make it difficult for users to tap accurately, especially on devices with smaller screens. \label{app:C17}
            \item Limited question visibility: The question text does not have a maximum line limit, which may cause issues if the question is too long, leading to the text being cut off or the layout looking cluttered. \label{app:C18}
            \item Ambiguous progress bar: The progress bar does not have a label or any indication of what it represents, which could lead to confusion about the user's progress within the quiz. \label{app:C19}
            \item Color contrast issues: The colors used for the “Next” button (yellow text on a yellow background) have low contrast, making it difficult to read, especially for users with visual impairments. \label{app:C20}
            \item No indication of the selected answer: There is no persistent indication of which answer was selected by the user after the answer is revealed, which can be problematic if the user wants to review their choice before moving on. \label{app:C21}
            \item Lack of confirmation before moving to the next question: There is no mechanism to confirm the user's choice before automatically advancing to the next question, which may lead to accidental selections. \label{app:C22}
        \end{enumerate}
    \item \textbf{Score View}
        \begin{enumerate}[resume, label=C\arabic*, ref=C\arabic*]
            \item The color contrast between the yellow buttons and the white text may not be sufficient for all users, particularly those with visual impairments, which can lead to readability issues. \label{app:C23}
            \item Lack of clear visual hierarchy could make it difficult for users to quickly understand the results; the results text and the buttons seem to have similar visual prominence. \label{app:C24}
            \item The buttons' labels “Try again!” and “Back” may not be descriptive enough to indicate their function clearly, which can lead to confusion about what action will be taken. \label{app:C25}
            \item There appears to be no indication of a correct answer review option, which might be beneficial for users who want to learn from their mistakes. \label{app:C26}
            \item The feedback provided to the user does not appear to be personalized beyond the number of correct answers; more detailed feedback could enhance the learning experience. \label{app:C27}
            \item There is no visual feedback on user interactions; for example, the buttons do not change appearance when tapped, which could leave users uncertain if their input has been registered. \label{app:C28}
            \item The spacing between the “Try again!” and “Back” buttons may not be enough to prevent accidental presses, which could lead to frustration. \label{app:C29}
            \item The fixed width and height of the buttons may not accommodate longer text for localization, potentially resulting in text clipping or layout issues in different languages. \label{app:C30}
        \end{enumerate}

\section*{To-Do App}
    \item \textbf{List View}
        \begin{enumerate}[resume, label=C\arabic*, ref=C\arabic*]
            \item Inconsistent visual feedback on task completion: The checkboxes provide a visual cue when a task is completed, but the accompanying text does not change its appearance, which could be improved by altering the text style to indicate completion more clearly. \label{app:C31}
            \item Lack of clear separation between tasks: The tasks are listed without clear dividers or spacing, which can make it difficult to distinguish individual tasks in longer lists. \label{app:C32}
            \item Ambiguous delete functionality: The delete icon's function in the navigation bar is not clear, and could be improved with a text label or a more descriptive icon. \label{app:C33}
            \item Missing visual hierarchy in list items: The task title and due date are styled similarly, making it hard to identify the most important information quickly. \label{app:C34}
            \item No indication of swipe actions: If swipe actions are available, they should be indicated visually to educate users. \label{app:C35}
            \item Redundant navigation links for completed tasks: Navigation elements for completed tasks could mislead users, and should be removed or visually distinguished. \label{app:C36}
            \item Unclear priority indication: The meaning of colored checkboxes as priority indicators is not obvious without a legend or explanation. \label{app:C37}
            \item Overloaded navigation bar: Having add and delete actions in the same place can be overwhelming; separating these actions can reduce cognitive load. \label{app:C38}
            \item Inconsistent interaction patterns: The use of toggles for marking tasks as completed may not align with other interaction patterns in the app. \label{app:C39}
            \item Lack of feedback when adding or removing tasks: No visual or auditory feedback is provided, which could leave users unsure of the action's success. \label{app:C40}
        \end{enumerate}
    \item \textbf{Item Detail View}
        \begin{enumerate}[resume, label=C\arabic*, ref=C\arabic*]
            \item Incomplete visual feedback for priority selection: No clear indicator like a checkmark or highlight to show the selected priority level when the picker is collapsed. \label{app:C41}
            \item Inconsistent navigation title: The title says “Edit task” which might be misleading for new tasks; it should change based on the action context. \label{app:C42}
            \item No visual separation between date picker and remove button: These elements are placed next to each other without clear visual separation, possibly causing confusion. \label{app:C43}
            \item No indication of mandatory fields: Fields like the title, which are required, are not marked, potentially leading to incomplete task saving attempts. \label{app:C44}
            \item Lack of confirmation or undo functionality for removing reminders: This can lead to accidental data loss. \label{app:C45}
            \item No clear error messaging for invalid due dates: The alert shown does not specify why the date is invalid or guide the user to correct it. \label{app:C46}
            \item Inconsistent button states: The “Done” button is disabled if the title is empty, but no similar constraints exist for other critical fields. \label{app:C47}
            \item No feedback on successful task creation or update: There is no visual confirmation that the task has been successfully saved. \label{app:C48}
            \item Limited accessibility support: UI elements lack descriptive labels for accessibility tools, hindering usability for visually impaired users. \label{app:C49}
        \end{enumerate}
\end{enumerate}

\section*{Interview Questions} \label{appendix:interview}
\begin{enumerate}[label=D\arabic*, ref=D\arabic*]
    \item Can you describe your team? \label{app:D1}
    \item What is your approach to combining UX with your development workflow? \label{app:D2}
    \item Did you face any challenges in combining UX into your development workflow? \label{app:D3}
    \item Did you conduct any usability evaluations for your app? \label{app:D4}
\end{enumerate}
    
    \textbf{Demonstration of UX-LLM on one of their app's view.}
    
\begin{enumerate}[resume, label=D\arabic*, ref=D\arabic*]
    \item On a scale of 1 to 4, how would you rate your initial impression of UX-LLM when it comes to improving the quality of an app's user experience? \label{app:D5}
    \item Based on what you have seen, what are your initial thoughts about UX-LLM? \label{app:D6}
\end{enumerate}

    \textbf{Presentation of usability issues generated by UX-LLM on the seven main views of their app.}
    
\begin{enumerate}[resume, label=D\arabic*, ref=D\arabic*]
    \item On a scale of 1 to 4, how useful do you find these usability issues? \label{app:D7}
    \item What are your initial thoughts about the usability issues identified by UX-LLM? \label{app:D8}
    \item How would you compare the filtered vs. unfiltered usability issues? \label{app:D9}
    \item On a scale of 1 to 4, how easy do you think it would be to integrate UX-LLM into your existing development workflow? \label{app:D10}
    \item How would you go about integrating UX-LLM into your workflow, and what factors do you think would make this easy or challenging? \label{app:D11}
    \item In general, what did you like about UX-LLM? \label{app:D12}
    \item Do you see any limitations or concerns with using UX-LLM for usability evaluations? \label{app:D13}
    \item What improvements or additional features would you suggest for UX-LLM? \label{app:D14}
    \item Do you have any other comments or thoughts you would like to share about UX in app development or about UX-LLM? \label{app:D15}
\end{enumerate}

\end{document}